\documentclass[12pt]{extarticle}
\usepackage{amsmath}
\usepackage{amssymb}
\usepackage{stmaryrd}
\usepackage{color}
\usepackage{ wasysym }
\usepackage{color}
\usepackage{enumitem}
\usepackage{graphicx}

\usepackage{soul}

   \newtheorem{theorem}{Theorem}[section]
   \newtheorem{lemma}[theorem]{Lemma}
   \newtheorem{proposition}[theorem]{Proposition}
   \newtheorem{corollary}[theorem]{Corollary}

\newcommand{\ben}{\begin{enumerate}}
\newcommand{\een}{\end{enumerate}}

\newcommand{\bt}{\begin{theorem}}
\newcommand{\et}{\end{theorem}}
\newcommand{\bl}{\begin{lemma}}
\newcommand{\el}{\end{lemma}}
\newcommand{\bc}{\begin{corollary}}
\newcommand{\ec}{\end{corollary}}
\newcommand{\bp}{\begin{proposition}}
\newcommand{\ep}{\end{proposition}}
\newcommand{\br}{\begin{remark}}
\newcommand{\er}{\end{remark}}

\newcommand{\bpf}{\begin{proof}}
\newcommand{\epf}{\end{proof}}

\newcommand{\be}{\begin{equation}} 
\newcommand{\ee}{\end{equation}}
\newcommand{\beq}{\begin{eqnarray}}
\newcommand{\eeq}{\end{eqnarray}}
\newcommand{\ba}{\begin{array}}
\newcommand{\ea}{\end{array}}
\newcommand{\bi}{\begin{itemize}}
\newcommand{\ei}{\end{itemize}}

\newcommand{\comm}[1]{}

\newcommand{\rar}{\rightarrow}

\newcommand \qed {\hskip 6pt\vrule height6pt width5pt depth1pt \bigskip}
\renewcommand{\theequation}{\arabic{section}.\arabic{equation}}

\newcommand{\bbC}{{\mathbb C}}

\newcommand{\bbR}{{\mathbb R}}

\newcommand{\bbZ}{{\mathbb Z}}

\newcommand{\psil}{{\psi_1}}
\newcommand{\psir}{{\psi_2}}

\newfont{\msbm}{msbm10 scaled\magstep1}
\newfont{\msbms}{msbm7 scaled\magstep1} 

   \newenvironment{proof}[1][Proof]{\begin{trivlist}
   \item[\hskip \labelsep {\bfseries #1}]}{\end{trivlist}}
   \newenvironment{definition}[1][Definition]{\begin{trivlist}
   \item[\hskip \labelsep {\bfseries #1}]}{\end{trivlist}}
   
   \newenvironment{remark}[1][Remark]{\begin{trivlist}
   \item[\hskip \labelsep {\bfseries #1}]}{\end{trivlist}}

   \comm{\newcommand{\qed}{\nobreak \ifvmode \relax \else
      \ifdim\lastskip<1.5em \hskip-\lastskip
      \hskip1.5em plus0em minus0.5em \fi \nobreak
      \vrule height0.75em width0.5em depth0.25em\fi}}

 \numberwithin{equation}{section}

\newcommand{\supp}{\mathop{\rm supp}}
\renewcommand{\Im}{\mathop{\rm Im}}
\renewcommand{\Re}{\mathop{\rm Re}}
\newcommand{\Log}{{\rm Log}}
\newcommand{\twovec}[2]{\begin{bmatrix}#1\\#2\\\end{bmatrix}}
\newcommand{\GF}{\textrm Green's function}
\def\DMO{\DeclareMathOperator}
\DMO{\ad}{ad}
\newcommand\numberthis{\addtocounter{equation}{1}\tag{\theequation}}

\begin{document}

\title{Resonances - lost and found}
\author{Richard Froese and Ira Herbst}

\maketitle
\begin{abstract}
We consider the large $L$  limit of one dimensional Schr{\"o}dinger operators
$
H_L=-d^2/dx^2 + V_1(x) +  V_{2,L}(x)
$
in  two cases:  when $V_{2,L}(x)=V_2(x-L)$ and  when $V_{2,L}(x)=e^{-cL}\delta(x-L)$. This is motivated by some recent work of Herbst and Mavi where $V_{2,L}$ is replaced by a Dirichlet boundary condition at $L$.  
The Hamiltonian $H_L$ converges to $H = -d^2/dx^2 + V_1(x)$ as $L\to \infty$ in the strong resolvent sense (and even in the norm resolvent sense for our second case). However, most of the resonances of $H_L$ do not converge to those of $H$. Instead, they crowd together and converge onto a horizontal line: the real axis in our first case and the line $\Im(k)=-c/2$ in our second case. In the region below the horizontal line resonances of $H_L$ converge to the reflectionless points of $H$ and to those of $-d^2/dx^2 + V_2(x)$.  It is only in the region between the real axis and the horizontal line (empty in our first case) that resonances of $H_L$ converge to resonances of $H$. Although the resonances of $H$ may not be close to any resonance of $H_L$ we show that they still influence the time evolution under $H_L$ for a long time when $L$ is large.
\end{abstract}

\tableofcontents

\section{Introduction}

A compactly supported potential $V(x)$ in one dimension with high barriers will give rise to shape resonances close to the real axis. These resonance energies $z_j$ and their associated resonant states $\phi_j$  determine the time evolution under $H=-d^2/dx^2 + V$ of an initial state  $\phi$ whose energy  and position are suitably localized near $\Re z_i$ and $\supp(V)$.  
Roughly speaking, for $t\ge 0$ we expect such a $\phi$ to obey
\[
\langle \phi, e^{-itH}\phi \rangle \sim \sum_{j=1}^n |\langle \phi, \phi_j \rangle|^2 e^{-itz_j}.
\]
What happens when a Dirichlet condition is imposed at $x=L$?
In a recent paper, Herbst and Mavi [HM] answered this question  when $V$ is a sum of two delta functions. Call the Hamiltonian with the additional Dirichlet condition $H_L$. It seems reasonable to expect that for a long time (roughly the time it takes $\phi$ to travel a distance $L$ under the free time evolution) the delta function should have little effect. In this time interval, the  time evolution under $H_L$ should be close to that under $H$ and so continue to be determined by the shape resonances. This turns out to be true. Given this,  it is reasonable to expect that the resonances of $H_L$ should converge to those of $H$ as $L\rar\infty$. This turns out to be false in a spectacular way. Instead of converging to the resonances of $H$, the resonances of $H_L$ cluster together and move toward the real axis, leaving every compact set in the lower half plane. 

The present work began with the observation that by modifying $V$ slightly, it is possible that {\it some} of the resonances of $H_L$ may not move towards the real axis, but instead  converge to points in the open lower half plane. However, these points are not  resonances of $H$! Moreover,  numerical experiments suggested that there are special points on the real axis {which} seem to repel the approaching resonances. 

To begin we will consider one dimensional Hamiltonians of the form $H_L=-d^2/dx^2 + V_1(x) + V_2(x-L)$, where $V_1$ and $V_2$ are compactly supported.
We give asymptotic formulas for the positions of the resonances of $H_L$ as $L\rar\infty$. The special points, both those in the lower half plane where resonances may approach and those on the real axis where the imaginary parts are repelled, turn out to be energies at which {a certain} reflection coefficient of the (meromorphic continuation of the) scattering matrix for either $V_1$ or $V_2$  vanishes. We will call these reflectionless points. (That $V_1$ and $V_2$ play a symmetrical role here is perhaps not so surprising considering that $H_L=-d^2/dx^2 + V_1(x) + V_2(x-L)$ {has the same resonances as} $-d^2/dx^2 + V_1(x+L) + V_2(x)$.) We will prove three asymptotic formulas, two for resonances near a point in $\bbR$, depending on whether the point is reflectionless, and one for resonances near a reflectionless point in the lower half plane. 

Next we examine what happens when we try to improve the convergence of $H_L$ to $H$ by adding an exponentially decreasing coupling constant. We consider $H_L=-d^2/dx^2 + V_1(x) +e^{-cL} \delta(x-L)$  and give asymptotic formulas for resonance positions near $k_0$ in six cases: three for  $k_0$ on the line $\Im(k_0)=-c/2$ depending on whether $k_0$ is a resonance for $H_1$, a reflectionless point for $H_1$ or neither, one for $k_0$ a reflectionless point for $V_1$ below the line $\Im(k)=-c/2$, one for $k_0$ a resonance for $V_1$ above the line $\Im(k)=-c/2$. In addition, {there may be} one resonance of $H_L$ converging to zero.  {As we will explain, similar phenomena occur when $e^{-cL} \delta(x-L)$ is replaced with $\mu(L) V(x-L)$ when $\mu(L) = e^{-cL}$ and $V$ is continuous with compact support.  In fact the exponentially decreasing coupling is the critical rate of decrease.  If it is slower than exponential the resonances of $H$ all disappear while if it is faster than any exponential, the resonances of $H_L$ converge to those of $H$.}

{For movies showing the large $L$ behavior of resonances of $H_L$ for both constant and decreasing coupling see the accompanying files resmovies.html, resc00.mp4 and resc16.mp4.}

 
Finally, we show that for long times (depending on $L$) the time evolution of a state $\phi$ under $H$ is close to the time evolution under $H_L$.

The lost resonances in the title of this paper refer to the original resonances  {of $H$} which disappear and are not evident in the resonance set of $H_L$ as $L\rightarrow\infty$, even though they show  up in the description of the time evolution of resonant states under $H_L$.
The found resonances are the resonances which are crowding together and moving toward the real axis {(or the line $\text {Im} (k) = -c/2$)} as $L$ becomes large.


\section{Hamiltonian and resonances}

We will consider Hamiltonians defined as self-adjoint realizations of $H=-d^2/dx^2 + V$ acting in $L^2(\bbR )$, where the potential $V$ has compact support.  For simplicity, and since our goal is not to treat the most general potentials, assume that 
$$V(x)=V_0(x) + \sum_{i=1}^N \alpha_i \delta(x-x_i),$$ 
where $V_0$ is continuous with compact support. 

The resolvent $(H-z)^{-1}$ for $z\in \bbC \backslash ([0,\infty) \cup \{\hbox{eigenvalues}\})$ has an integral kernel given by the \GF
\[
G(x,y;k) = \frac{1}{W(k)}\begin{cases} \psil(x;k)\psir(y;k) &x\le y\\\psil(y;k) \psir(x;k) &y\le x\\\end{cases}
\]
where $\Im k > 0$, $k^2=z$ and $\psil$ and $\psir$  are solutions  to
 \begin{equation}\label{eqn:SSE}
 -\psi'' + V(x)\psi = k^2\psi,
 \end{equation}
 satisfying $\psil(x;k)=e^{-ikx}$ for $x\le\inf\supp(V)$ and $\psir(x;k)=e^{ikx}$ for $x\ge\sup\supp(V)$. The Wronskian 
 \[
W(k) = \det\left(\twovec{\psil(x,k) & \psir(x,k) }{\psil'(x,k)  & \psir'(x,k) }\right)
\] 
does not depend on $x$. 

When the potential includes a non-zero  delta function term $\sum_{i=1}^N \alpha_i \delta(x-x_i)$ then  \eqref{eqn:SSE} is interpreted to mean that $\psi(x)$ is a  classical $C^2$ solution in the open intervals between the points $x_i$, while at the points $x_i$, $\psi(x)$ is continuous and $\psi'(x)$ has left and right limits that are related by
\[
\psi'(x_i+) = \psi'(x_i-)+\alpha_i\psi(x_i).
\]
This condition ensures that $\psi$ is (locally) in the domain of $H$, {(that is $\phi \psi $ is in the domain of $H$ for all $\phi \in C_0^{\infty}(\mathbb R))$}. (See [AR] for the meaning of this equation when $V$ is a finite Borel measure).

The solutions $\psil(x;k)$ and $\psir(x;k)$ are defined for any complex $k$. For fixed $x$ they are  entire functions of $k$.  The derivatives $\psil'(x;k)$ and $\psir'(x;k)$ for fixed $x$ ({or} the right and left limits when $x=x_i$) are also entire functions of $k$. 
Given these analyticity properties, we see that the \GF\  has a meromorphic continuation to all of $\bbC$ with poles at the zeros of $W(k)$. 

The zeros of $W(k)$ can be characterised as those $k$ values for which $\psil(x;k)$ is a multiple of $\psir(x;k)$.
When this happens for $\Im(k)>0$ the solution $\psil$ is an eigenfunction so that $E=k^2$ is an eigenvalue of $H$.
This can only occur at points $k$ on the positive imaginary axis.  

\begin{definition} The {\it resonances} of $H$ are poles in the continuation of the \GF\ in $k$ to the lower half plane. Equivalently, they are zeros of $W(k)$ in the lower half plane. 
\end{definition}
Another characterisation of resonances is that they are values of $k$ such that for some constants $a$ and $b$,  $\psil(x;k) = a e^{ikx}$ for large positive $x$ and $\psir(x;k) = b e^{-ikx}$ for large negative $x$.
\begin{definition}
The {\it reflectionless points} are values of $k$ such that for some constants $c$ and $d$, $\psil(x;k) = c e^{-ikx}$ for large positive $x$  {or} $\psir(x;k) = d e^{ikx}$ for large negative $x$. They are the values of $k$ where  {a} reflection coefficient in the {(meromorphically continued)} scattering matrix vanishes. 
\end{definition}

With these definitions resonances and reflectionless points are $k$ values.  The corresponding complex energies are at $z=k^2$ on the second sheet obtained by analytic continuation across the positive real axis. 

Resonances are values of $k$ for which $\psil(x,k)$ is a multiple of $\psir(x,k)$. 
We now analyze this condition when $V(x)=V_1(x)+V_2(x-L)$ with $\supp(V_1)\subseteq [x_0,0]$ and $\supp(V_2)\subseteq [0,x_1]$. For the moment we also assume that $k\ne 0$. Since $V(x)=0$ for $x\in(0,L)$, any solution $\psi(x;k)$ of  \eqref{eqn:SSE} is a linear combination $\psi(x;k)=c_1 e^{-ikx} + c_2 e^{-ikx}$ for $x$ in this interval. This implies that for any solution $\psi(x;k)$
\[
\twovec{\psi(L;k) }{\psi'(L-;k) } = \twovec{\cos(kL)&\sin(kL)/k}{-k\sin(kL)&\cos(kL)}\twovec{\psi(0;k) }{\psi'(0+;k) }.
\]
Since $k\ne 0$, $\twovec{\phantom{-}ik&1}{-ik&1}$ is invertible and we have
\[
\twovec{\phantom{-}ik&1}{-ik&1}\twovec{\cos(kL)&\sin(kL)/k}{-k\sin(kL)&\cos(kL)}\twovec{\phantom{-}ik&1}{-ik&1}^{-1}
=\twovec{e^{ikL}&0}{0&e^{-ikL}}
\]
so that
\begin{equation}\label{eqn:PP}
\twovec{\phantom{-}ik\psi(L;k) +\psi'(L-;k)}{-ik\psi(L;k) +\psi'(L-;k)}=\twovec{e^{ikL}&0}{0&e^{-ikL}}\twovec{\phantom{-}ik\psi(0;k) +\psi'(0+;k)}{-ik\psi(0;k) +\psi'(0+;k)}.
\end{equation}
Now suppose that $k$ is a non-zero resonance.  We apply \eqref{eqn:PP} to $\psil$ and use the fact that $\psir$ is a multiple of $\psil$. This yields
\begin{equation}\label{eqn:R}
e^{2ikL} = f(k)
\end{equation}
where 
\begin{equation}\label{eqn:f}
f  = f_1 f_2, \quad
f_1(k)  =\left(\frac{-ik\psil(0;k) +\psil'(0+;k)}{\phantom{-}ik\psil(0;k) +\psil'(0+;k)}\right), \quad
f_2(k) =\left(\frac{\phantom{-}ik\psir(L,k) +\psir'(L-;k)}{-ik\psir(L,k) +\psir'(L-;k)}\right).
\end{equation} 
On the other hand, if \eqref{eqn:R} holds with $f=f_1f_2$ given by \eqref{eqn:f}, then \eqref{eqn:PP} holds with $\psir$ on the right and $\psil$ on the left, up to a non-zero multiple. Equation \eqref{eqn:PP} also holds with $\psil$ on both sides. Thus we find that $\psil$ and $\psir$ satisfy the same initial condition up to a multiple. This implies that $\psil(x;k)$ is a multiple of $\psir(x;k)$ which means that $k$ is a resonance. Thus the non-zero resonances for $V_1(x)+V_2(x-L)$ can be characterized as solutions $k$ in the lower half plane of \eqref{eqn:R}, where $f$ is given by \eqref{eqn:f}. 

Notice that $f(0)=1$ provided $\psi_1' (0+;0)$ and $\psi_2' (L-;0)$ are not zero. This is generically true when{ $V_1$ and $V_2$  are} not zero. Thus, $k=0$ is generically a solution to \eqref{eqn:R}, even if the \GF\ does not have a pole at $0$.
For example, if $V_1(x)=V_2(x)=\alpha\delta(x)$ then $f(k)=(\alpha-2ik)^2/\alpha^2$ while $W(k) = 2(\alpha -ik) + i\alpha^2(1-e^{2ikL})/2k$. Here $W(0)=L\alpha^2 + 2\alpha$ is non-zero, so $k=0$ is not a pole of the \GF. However $f(0)=1$ in this example so that $e^{2i0L}=f(0)$.

Notice that $\psil(0;k) = \psil(0;k;V_1)$ (the third argument denotes the relevant potential) while $\psir(L;k)=\psir(0;k;V_2)$. This implies that $f(k)$ is independent of $L$. Moreover $f(k)$ has a pole if $\psir'(L-;k;V_2)=ik\psir(L;k;V_2)$ or $\psil'(0+;k;V_1)=-ik\psil(0;k;V_1)$.   The first condition implies that the solution $\psir(x;k;V_2)$ is a multiple of $e^{ikx}$ both for $x<0$ and $x>x_1$. Thus $k$ is a reflectionless point for $V_2$. Similarly, the second condition holds if $\psil(x;k;V_1)$ is a multiple of $e^{-ikx}$ both for $x<x_0$ and $x>0$, that is, if $k$ is a reflectionless point for $V_1$. This shows that the poles of $f(k)$ occur at the reflectionless points of either $V_1$ or $V_2$. Similarly, the zeros of $f(k)$ occur at the resonances of either $V_1$ of $V_2$. {However if $k$ is a resonance of $V_1$ and at the same time a reflectionless point of $V_2$ (or vice versa), $k$ will in general be neither a pole nor a zero of $f$.}

We will need one more property of $f(k)$, namely that $|f(k)| > 1$ when {$k \in \mathbb R$}. To see this we define
$$
w_1(x;k)=\psil'(x;k)/\psil(x;k),
$$
which should be thought of as an affine co-ordinate for the complex direction of $\twovec{\psil(x;k) }{\psil'(x,k) }$ in $\bbC^2$. We allow $w_1=\infty$ to include the direction where $\psil=0$. We have $w_1(x;k) = -ik$ for $x\le x_0$. For $x$ between the points $x_i$,
$w_1(x;k)$ solves the Ricatti equation
$$
w_1'(x;k) = 	V(x) - k^2 - w_1^2(x;k).
$$ 
At the points $x_i$, $w_1(x;k)$ jumps to the right by $\alpha_i$ with the convention that $\infty+\alpha_i=\infty$. For a better understanding of what happens near a point where $w_1=\infty$ define $u_1(x;k)=1/w_1(x;k)$. Then $u_1$ satisfies the equation $u_1' = 1+u_1^2(k^2-V(x))$ which is well defined near $u_1=0$. When $k$ and $w_1$ are both real, the right side of the Ricatti equation is real.   Moreover, the jumps at the points $x_i$ are also in the real direction. This implies that if  $w_1(x;k)$ is real for some $x$ it must be real for all $x$ and cannot satisfy the initial condition.  Therefore for $k > 0$ $w_1(x;k)$ must stay in the lower half plane. For $k >0$ the fractional linear transformation $w\mapsto (w-ik)/(w+ik)$ maps the lower half plane to the exterior of the unit disk. Thus for $k >0$,
$$
\left|\frac{w_1(0;k)-ik}{w_1(0;k)+ik}\right| = \left|\frac{-ik\psil(0;k) +\psil'(0+;k)}{ik\psil(0;k) +\psil'(0+;k)}
\right| >1
$$
A similar argument using $w_2(x;k)=\psir'(x;k)/\psir(x;k)$ shows
$$
\left|
\frac{ik\psir(L,k) +\psir'(L-;k)}{-ik\psir(L,k) +\psir'(L-;k)}
\right|>1
$$
as well.  We see that $\bar{f}(k) = f(-\bar{k})$ which implies $|f(k)| > 1$ for $k < 0$.

The inequality $|f(k)|>1$ for  $k  \in \mathbb{R}\setminus \{0\}$ 
implies that \eqref{eqn:R}  has no real non-zero solutions since $|e^{2iLk}|=1$ for these $k$ values.

Summarizing, we have proved the following proposition.
\begin{proposition} Let $f(k)$ be defined by \eqref{eqn:f}. A {nonzero} complex number $k$ solves \eqref{eqn:R}
if and only if
\begin{enumerate}
\item $k$ is on the positive imaginary axis and $k^2$ is an eigenvalue, or 
\item $k$ is a resonance in the open lower half plane. 
\end{enumerate}
The function $f(k)$ is a meromorphic function with poles at the reflectionless points of either $V_1$ or $V_2$ and zeros at the resonances of either $V_1$ or $V_2$. When {$k \in \bbR\backslash\{0\}$}, $  |f(k)| >1$. We have $\bar{f}(k) = f(-\bar{k})$ and thus if $k$ is a resonance then so is $-\bar k$.
\end{proposition}

\section{Resonances as $L\rightarrow\infty$ when $V_{2,L}(x) = V_2(x-L)$}\label{no coupling constant}

In this section we determine the asymptotic positions of the resonances of $H_L=-d^2/dx^2 + V_1(x) + V_2(x-L)$ for large $L$.

Let ${\cal R}(L)$ denote the set of solutions to \eqref{eqn:R} where $f(k)$ is the function given by \eqref{eqn:f} and let $\bbC_-=\{k\in\bbC:\Im(k)<0\}$. Then ${\cal R}(L)\cap \bbC_-$ is the set of resonances. However,  {generically,} ${\cal R}(L)$  contains $k=0$ even if the \GF\ has no pole there.
We will locate the asymptotic positions of the points in ${\cal R}(L)$. A consequence of our estimates is:

\begin{theorem} \label{thm:counting}

Let  $N(L)=\#\Big({\cal R}(L)\cap \big([a,b]\times [-ic,0]\big)\Big)$ for $0\le a<b<\infty$ and $c>0$.  We then have

$$\lim _{L \to \infty} N(L)/L = (b-a)/\pi$$
If $[a,b]$ contains no poles of $f$ then there is a constant $c_1> 0$ so that the number of resonances in $[a,b] \times [ -ic_2,-ic_1 L^{-1} ]$ is bounded uniformly in $L$ as $L \to \infty$ for any $c_2 > 0$.  If $[a,b]$ contains a pole of $f$ there is a $c'_1> 0$ so that the number of resonances in $[a,b] \times [ -ic_2, -ic'_1 L^{-1}\log L]$ is bounded uniformly in $L$ as $L \to \infty$ for any $c_2 > 0$.  If $[a,b]$ contains no poles of $f$ of order higher than $p$ then we can take $c'_1 = p/2$.

\end{theorem}
To begin, we show that the sets
$$
U(a,A) = \{k\in\bbC_- : \Im(k) < -a \hbox{\ and\ }|f(k)| < A\}
$$
for $a,A>0$ are resonance free for $L$ sufficiently large. Recall that reflectionless points are poles of $f$.
So the sets $U(a,A)$ exclude a strip below the real axis together with some neighbourhoods of the reflectionless points.
\begin{proposition}\label{leaving}
There exists $L_0$ such that $U(a,A)\cap{\cal R}(L) = \emptyset$ when $L>L_0$.
\end{proposition}
\begin{proof} Comparing the moduli of the left and right sides of \eqref{eqn:R},
we find that any resonance in $U(a,A)$ satisfies
\[
e^{2La}<|e^{2iLk}| = |f(k)| < A,
\]
so that $L<\log(A)/(2a)$. This proves the proposition with $L_0=\log(A)/(2a)$.
\end{proof}

We will use the following notation.  The disk centered at $k_0$ with radius $r$ is 
$$B(k_0,r)=\{k\in\bbC: |k-k_0|<r\}.$$
For $f$ analytic near $k_0$ with $f(k_0)\ne 0$
$$
\Omega(f,k_0,\epsilon) 
=\left\{k\in\bbC : \left|1-\dfrac{f(k)}{f(k_0)}\right|<\epsilon\right\}.
$$

Now let us find all the resonances in a neighbourhood of $k_0$ in the closed {lower} half plane $\overline{\bbC}_-$ for $L$ large. Given proposition \ref{leaving} we need only consider $k_0\in\bbR$ and $k_0$ a pole of $f(k)$. We will distinguish three cases.

\bigskip
\noindent{\bf Case 1: $k_0\in\bbR$ and $f$ is analytic at $k_0$}\\

Let $\log$ be any branch of the logarithm {and $r>0$ such that $f(k) \ne 0$ for $k \in B(k_0,r)$. }Then 
$$
{\cal R}(L)   {\cap B(k_0,r)} =\left\{k  { \in B(k_0,r)}:k=\dfrac{1}{2iL}\big(\log(f(k)) + 2\pi ij\big) \hbox{\ for some\ } j\in\mathbb Z\right\}.
$$
We call $k_j$ an approximate resonance for $L$ near $k_0$ if
$$
k_j = \dfrac{1}{2iL}\big(\log(f(k_0)) + 2\pi ij\big)
$$
for some $j\in\bbZ$, and denote the set of these by ${\cal A}(k_0,L)$. Notice that $f(k_0)\ne 0$, since $k_0\in\bbR$ implies $|f(k_0)|\ge 1$.

\begin{proposition} \label{prop:nopole}
Suppose $f(k)$ is analytic near $k_0\in\bbR$.  Let $\epsilon \in(0,1/2)$. Then there exists $\delta>0$ and $L_0$ such that for all $L>L_0$
\begin{enumerate}[label=(\roman*)]
\item If $k_j\in B(k_0,\delta) \cap {\cal A}(k_0,L)$ then there is exactly one $k\in{\cal R}(L)$ with $|k-k_j|<\epsilon/L$.
\item If $k\in B(k_0,\delta) \cap {\cal R}(L)$ then there is a $k_j\in {\cal A}(k_0,L)$ with $|k-k_j|<\epsilon/L$.
\end{enumerate}
\end{proposition}
\begin{proof}
Let $\log$ be a branch of the logarithm such that $\log(f(k))$ is analytic for $k\in\Omega(f, k_0,\epsilon)$. This is possible because if $k\in\Omega(f, k_0,\epsilon)$ then $|f(k_0)-f(k)| < (1/2)|f(k_0)|$. Thus $f(k)$ lies in a disk of radius $|f(k_0)|/2$ centred at $f(k_0)$, and we can find a branch cut that misses this disk. {Choose $0 < \delta \le r/2 $ small enough so that $B(k_0, 2\delta) \subset \Omega(f, k_0,\epsilon)$}.  For $j\in\bbZ$, define
$$
\phi_j(k) = k - \dfrac{1}{2iL}\big(\log(f(k) + 2\pi ij\big)
$$
so that  {$k \in {\cal R}(L) \cap B(k_0, 2\delta)$} if and only if  {$k\in B(k_0,2\delta)$ and} $\phi_j(k)=0$ for some $j\in\bbZ$.
 Let $L_0=2/\delta$ and assume $L > L_0$. If $k_j$ is an approximate resonance for $L$ in $B(k_0,\delta)$ then $B(k_j,\epsilon/L)\subset B(k_0,2\delta)$ so both $\phi_j$ and $\xi_j(k) = k-k_j$ are analytic in a neighbourhood of  {$\overline{B(k_j,\epsilon/L)}$}. Clearly $|\xi_j(k)| = \epsilon/L$ for $k$ on the boundary of $B(k_j,\epsilon/L)$.
On the other hand
\begin{align*}
|\xi(k)-\phi_j(k)| &=\dfrac{1}{2L}\left|\log\left(\dfrac{f(k)}{f(k_0)}\right)\right|\\
&=\dfrac{1}{2L}\left|\log\left(1+\dfrac{f(k)}{f(k_0)} -1\right)\right|\\
&\le \dfrac{1}{2L}\dfrac{\left|\dfrac{f(k)}{f(k_0)} -1\right|}{1-\left|\dfrac{f(k)}{f(k_0)} -1\right|}\\
&< \dfrac{1}{2L}\dfrac{\epsilon}{1-\epsilon}\\
&<\epsilon/L.\\
\end{align*}
Here we used that $|\log(1+z)|\le |z|/(1-|z|)$ for $|z|<1$ and that $\epsilon<1/2$. Thus Rouch\'e's theorem implies that $\xi$ and $\phi_j$ have the same number of zeros in $B(k_j,\epsilon/L)$, namely one. This proves (i).

To prove (ii) we suppose that we have a resonance $k\in B(k_0,\delta)\cap{\cal R}(L)$. Then 
$$
k = \dfrac{1}{2iL}\big(\log(f(k) + 2\pi ij\big)
$$for some $j\in\bbZ$. For this $j$ define
$$
k_j= \dfrac{1}{2iL}\big(\log(f(k_0) + 2\pi ij\big).
$$
Then, since  $B(k_0,\delta)\subset\Omega(f, k_0,\epsilon)$. 
$$
|k-k_j| = \dfrac{1}{2L}\left|\log\left(\dfrac{f(k)}{f(k_0)}\right)\right|<\epsilon/L
$$
as before.
\end{proof}

\bigskip
\setlength{\fboxsep}{10pt}
\fbox{\begin{minipage}{40em}
We will illustrate the approximations with  $V_1 = \delta(x+1)+\delta(x)$ and $V_2=\delta(x)+\delta(x-1)$. Then
$$
f(k)=\frac{(-e^{2ik}+1-4ik-4k^2)^2}{(-e^{2ik}+1+2ik(-e^{2ik}-1))^2}.
$$
There are infinitely many poles on the real axis but none in $\bbC_-$. So all the resonances will be converging to the real axis. We pick a point that is not a pole, say $k_0=3$ (blue dot) and compute the the exact (red circle) and approximate (blue crosses) resonances. Here we have taken $L=30$.
\hskip 20pt

\begin{center}
\includegraphics[width=8cm]{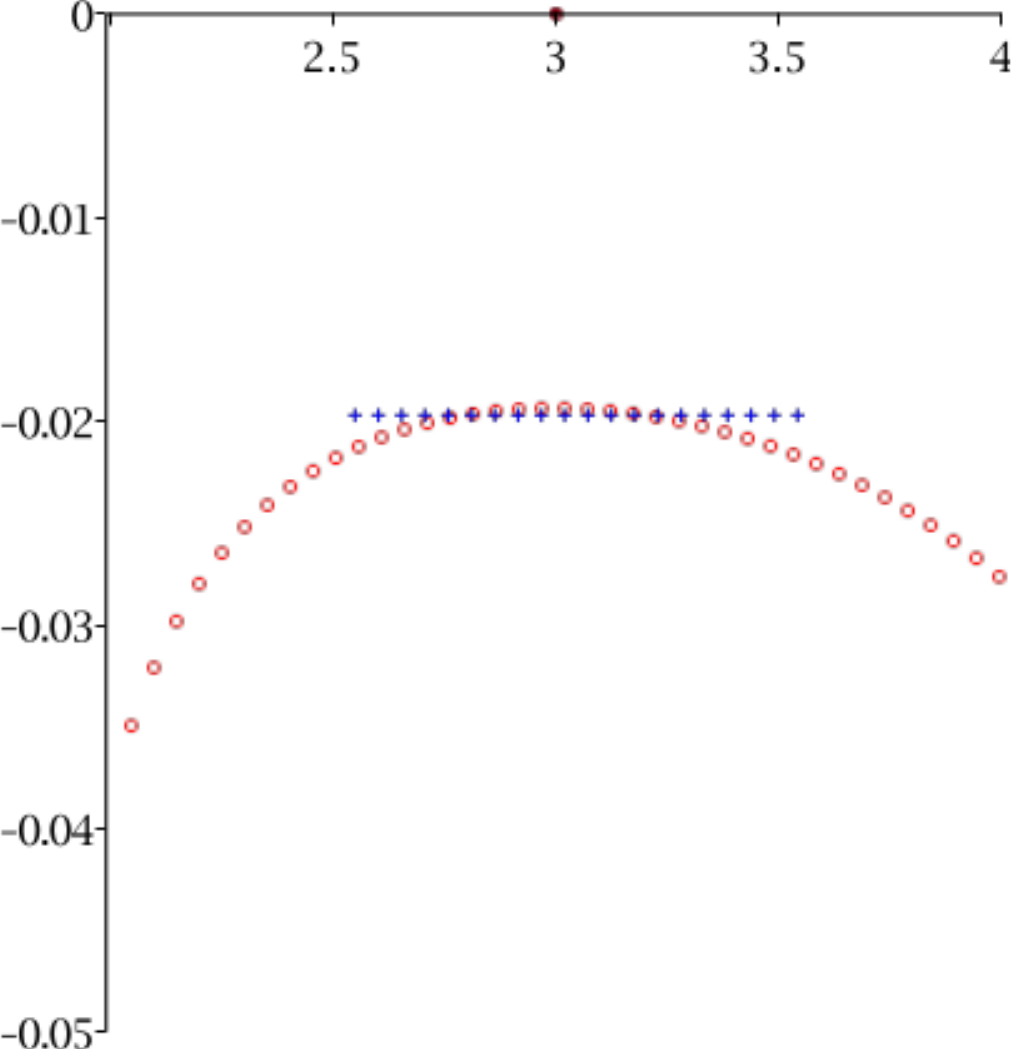}
\end{center}

\end{minipage}}

\bigskip\bigskip
\noindent{\bf Case 2: $k_0\in\bbR$ and $k_0$ is a pole of $f$}.

Write $f(k)=\dfrac{g(k)}{(k-k_0)^p}$ where $p$ is a positive integer, $g$ is analytic at $k_0$ and $g(k_0)\ne 0$. Then { for $k$ near $k_0$ ,}$k\in {\cal R}(L)$ whenever
$$
e^{2iLk}=\dfrac{g(k)}{(k-k_0)^p}
$$
or, equivalently,
\begin{equation}\label{eq:resp}
(k-k_0)^p e^{2iL(k-k_0)}=  e^{-2iLk_0}g(k)
\end{equation}
Choose a branch of $G(k)$ of $g^{1/p}(k)$ that is analytic in a neighbourhood of $k_0$. To simplify notation define
$\lambda = 2L/{p}$ and $\omega=e^{2\pi i /p}$.  Then \eqref{eq:resp} holds if and only if 
\begin{equation}\label{eq:resproot}
i\lambda(k-k_0) e^{i\lambda (k-k_0)}= i\lambda e^{-i\lambda k_0}G(k)\omega^l
\end{equation}
for some $l\in \{0,1,\ldots, p-1\}$. 
The solutions $z$ of $ze^z = w$ are (by definition) branches of the Lambert $W$ function. It follows from \eqref{eq:resproot} that $k$ is a resonance for  {$H_L$} if and only if
$$
k=k_0 + \frac{1}{i\lambda}W(A(k,k_0,l,\lambda))
$$
for some branch $W$ of the Lambert $W$ function, where 
$$
A(k,k_0,l,\lambda) = i\lambda e^{-i\lambda k_0}\omega^lG(k),
$$
We want to mention that recently in [Sa] the Lambert $W$ function was used to calculate shape resonances induced by two delta function barriers.
Let $W_j(z)$ for $j\in\bbZ$ be the branches of the Lambert $W$ function defined in [CGHJK]. Define approximate resonances
\begin{equation}\label{approxres2}
k_{j,l}=k_0 + \frac{1}{i\lambda}W_j(A(k_0,k_0,l,\lambda))
\end{equation}
for $j\in\bbZ$ and $l\in \{0,1,\ldots, p-1\}$.  We will denote the set of approximate resonances near $k_0$ by ${\cal B}(k_0,\lambda)$.

\begin{proposition}\label{asympt1}
Suppose $f(k)$ has a pole of order $p$ at $k_0\in\bbR$.  Let $\epsilon \in(0,1/2)$. Then there exists $\delta>0$ and $\lambda_0$ such that for all $\lambda>\lambda_0$
\begin{enumerate}[label=(\roman*)]
\item If $k_{j,l}\in B(k_0,\delta) \cap {\cal B}(k_0,\lambda)$ then there is exactly one  $k\in{\cal R}(L)$ with $|k-k_{j,l}|<\epsilon/\lambda$.
\item If $k\in B(k_0,\delta) \cap {\cal R}(L)$ then there is a $k_{j,l}\in {\cal B}(k_0,\lambda)$ with $|k-k_{j,l}|<\epsilon/\lambda$.
\end{enumerate}
\end{proposition}

Before proving this proposition we collect some facts about the branches of the Lambert $W$ function. For the branches $W_j$, the branch points are at $-1/e$ for $j=0$, at $0$ and $-1/e$ for $j=1,-1$ and at $0$ for $|j|>1$. The branch cuts are on the negative real axis. 
The branches $W_j$ have  expansions of the form
$$
W_j(z) = \Log_j(z) -\log(\Log_j(z)) + \sum_{k=0}^\infty\sum_{m=1}^\infty c_{k,m}\frac{\log(\Log_j(z))^m}{\Log_j(z)^{k+m}}
$$
where $\log$ is the principal branch and $\Log_j(z) = \log(z) + 2\pi ij$. These expansions are convergent for large $|z|$. For $j\ne 0$ the expansions converge near $z=0$ too. 
For any branch $W=W(z)$ we have $e^{2|W|} \ge |We^W| = |z|$ so
\begin{equation}\label{eq:Wlb}
|W(z)| \ge (1/2)\log|z|
\end{equation}
The derivative of any branch in its region of analyticity can be found by implicit differentiation. We obtain
\begin{equation}\label{eq:Wderiv}
W'(z) = \frac{W(z)}{(1+W(z))z}
\end{equation}

\begin{lemma}\label{Wbound}
Let $a_0\in\mathbb C$ and $\epsilon\in(0,1/2)$ with $(1-\epsilon/4)|a_0| > e^3$. Let $B$ denote the disk $B(a_0,\epsilon|a_0|/4)$. For any $j\in\mathbb Z$ there is a branch $W_*$ of the Lambert $W$ function analytic in $B$ such that $W_*(a_0)=W_j(a_0)$. For any $a\in B$ we have
\begin{equation}\label{West}
|W_*(a) - W_*(a_0)| < \epsilon.
\end{equation}
\end{lemma}
\begin{proof}
Any $a\in B$ satisfies $|a| \ge (1-\epsilon/4)|a_0| > e^3$. Thus $B$ does not contain the possible the branch points of the  $W_j$ (i.e., $0$ and $-1/e$). So the analytic continuation $W_*$ of $W_j$ near $a_0$ will be analytic and single valued in $B$. 
For $a\in B$, let $[a_0,a]$ denote the straight line path from $a_0$ to $a$. Then we may estimate
\begin{align*}
|W_*(a) - W_*(a_0)| &= \left|\int_{[a_0,a]}W'_*(z) dz\right|\\
&\le \sup_{z\in B}\left| W'(z)\right ||a-a_0|\\
&\le \sup_{z\in B} \frac{|W_*(z)|}{|1+W_*(z)||z|}\epsilon|a_0|/4\\
&\le \sup_{z\in B}\left(1 + \frac{1}{|W_*(z)|-1}\right)\frac{\epsilon|a_0|/4}{(1-\epsilon/4)|a_0|}\\
&\le \sup_{z\in B}\left(1 + \frac{1}{\log|z|/2-1}\right)\frac{\epsilon}{(4-\epsilon)}\\
\end{align*}
Now we use that $|z|>(1-\epsilon/4)|a_0|>e^3$ which implies $\log|z| > 3$ and $\epsilon<1/2$ to conclude
$$
|W_*(a) - W_*(a_0)| \le \frac{3\epsilon}{4-\epsilon} < \epsilon.
$$
\end{proof}

\begin{proof} (of Proposition \ref{asympt1})
Given $\epsilon\in (0,1/2)$, choose $\delta$ sufficiently small so that $G(k)$ is analytic in $B(k_0,2\delta)$ and $B(k_0,2\delta)\subset \Omega(G,k_0,\epsilon/4)$. Let $\lambda_0 = \max\{e^3/((1-\epsilon/4)|G(k_0)|), \epsilon/\delta\}$.
Assume that $k_{j,l}$ is an approximate resonance given by \eqref{approxres2} for some $j\in\mathbb Z$ and $l\in\{0,\ldots,p-1\}$ and that $k_{j,l}\in B(k_0,\delta)$. We must show that there exists a resonance in $k\in{\cal R}(L)$ where $L=p\lambda/2$ with $|k-k_{j,l}|<\epsilon/\lambda$ whenever $\lambda > \lambda_0$.

To simplify notation, denote $A(k,k_0,l,\lambda)$ by $A(k)$. Note that  $k\in \Omega(G,k_0,\epsilon/4)$ implies that $A(k)\in B(A(k_0),\epsilon |A(k_0)|/4)$. In addition, $|A(k_0)|=\lambda|G(k_0)|$ so that $\lambda > \lambda_0$ implies $(1-\epsilon/4)|A(k_0)| > e^3$. Thus we may apply Lemma \ref{Wbound} with $a_0=A(k_0)$ to conclude that 
$$
|W_*(A(k))-W_*(A(k_0))| < \epsilon
$$
for $k\in B(k_0,2\delta)$ and $\lambda > \lambda_0$.

Now define $\xi_{j,l}(k)=k-k_{j,l}$ and $\phi_{j,l}(k)=k-k_0-(i\lambda)^{-1}W_*(A(k))$. We wish to apply Rouch\'e's theorem on $B(k_{j,l},\epsilon/\lambda)$.
Since $k_{j,l}\in B(k_0,\delta)$ and $\epsilon/\lambda < \delta$ for $\lambda > \lambda_0$ it follows that for $\lambda > \lambda_0$, $B(k_{j,l},\epsilon/\lambda)\subseteq B(k_0,2\delta)$. On the boundary of $B(k_{j,l},\epsilon/\lambda)$ we have $|\xi_{j,l}(k)|=\epsilon/\lambda$. On the other hand, for all $k\in B(k_0,2\delta)$ we also have
\begin{align*}
|\xi_{j,l}(k)-\phi_{j,l}(k)| 
&= \frac{1}{\lambda}\left|W_*(A(k)) -W_j(A(k_0))\right| \\
&= \frac{1}{\lambda}\left|W_*(A(k)) -W_*(A(k_0))\right| \\
&< \epsilon/\lambda,\\		
\end{align*}
provided $\lambda > \lambda_0$. Then Rouch\'e's theorem says that there is exactly one solution of $\phi_{j,l}(k)=0$, that is, exactly one resonance, in 
$B(k_{j,l},\epsilon/\lambda)$. This proves (i).

To prove (ii) we note that every resonance can be written $k = k_0 + (i\lambda)^{-1}W_j(A(k))$ for some some $j\in\mathbb Z$ and $l\in\{0,\ldots,p-1\}$.
If we know that $k\in B(k_0,\delta)$, with $\delta$ chosen as above, then for $\lambda > \lambda_0$ we will have $|W_*(A(k))-W_*(A(k_0))| < \epsilon$ as before. This implies that if we let $k_{j,l}$ be the approximate resonance $k_{j,l}=k_0 + (i\lambda)^{-1}W_j(A(k_0))$, then $|k-k_{j,l}| < \epsilon/\lambda$.

\end{proof}

\bigskip
\setlength{\fboxsep}{10pt}
\fbox{\begin{minipage}{40em}
Here are the computed resonances (red circles) and approximate resonances ($l=0$ blue and $l=1$ green) near the double pole $k_0\sim 4.81584231784594$.
\begin{center}
\includegraphics[width=10cm]{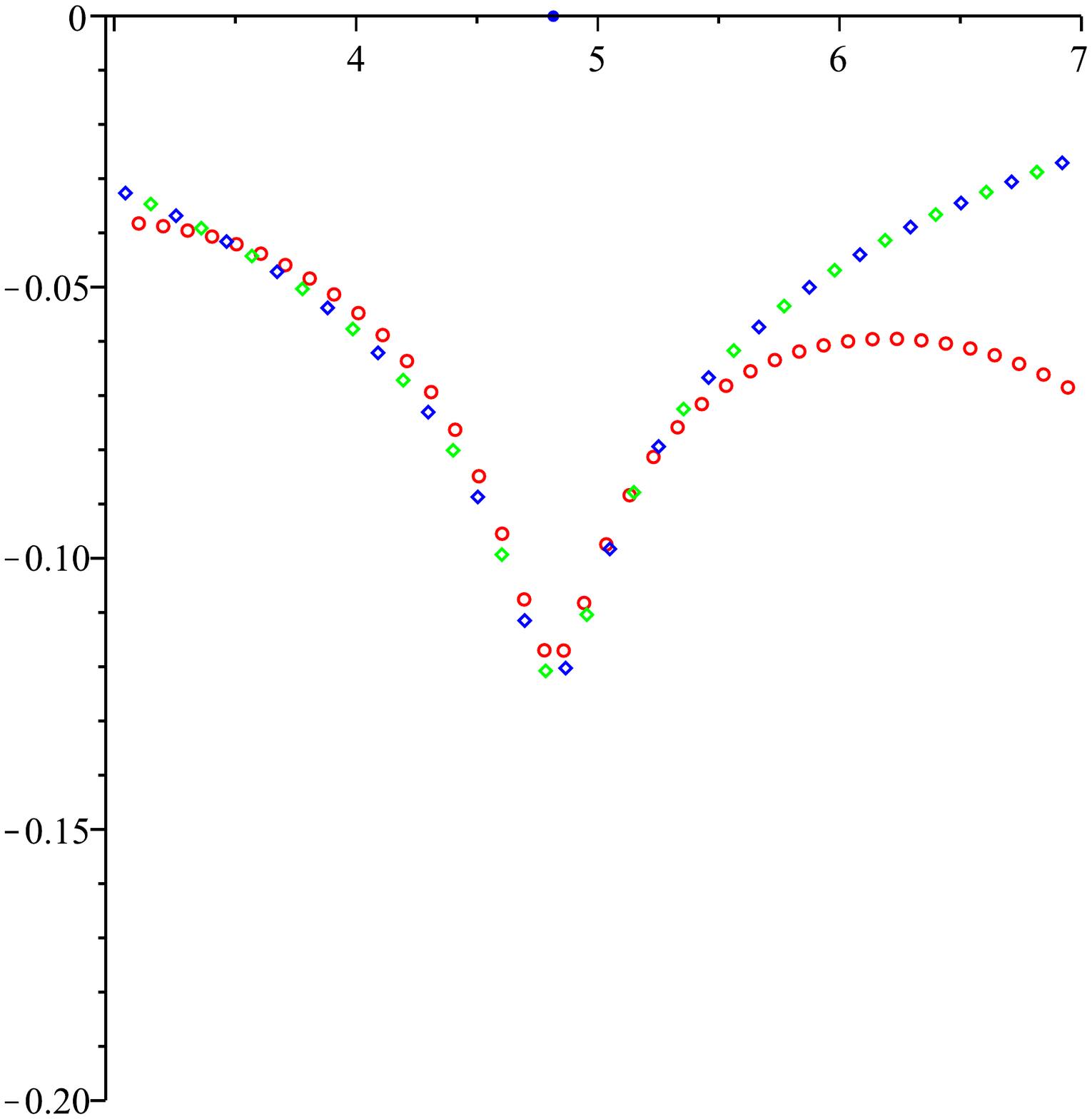}
\end{center}
\end{minipage}}
\bigskip\bigskip

We can make a further approximation of the approximate resonances that does not involve the Lambert $W$ function. Using the first two terms of the expansion for $W_j$ gives 
$$
k_{j,l} \sim k_0 + \frac{1}{i\lambda}\Big(\log(A_0) + 2\pi i j + \log(\log(A_0) + 2\pi i j)\Big)
$$
to an accuracy of $O\left(\dfrac{\log\log(\lambda)}{\lambda\log(\lambda)}\right)$. Here $A_0=A(k_0,k_0,l,\lambda)=i\lambda e^{-i\lambda k_0}\omega^lG(k_0)$ as before.
For $x\in\bbR$, let $[x]$ denote the integer contained in $(x-1/2,x+1/2]$ define $\Phi(k_0,l,\lambda)\in(-1/2,1/2]$ as
$$
\Phi(k_0,l,\lambda)=\frac{-\lambda k_0 + \varphi_0}{2\pi} + \frac{l}{p} + \frac{1}{4} - \left[\frac{-\lambda k_0 + \varphi_0}{2\pi} + \frac{l}{p} + \frac{1}{4} \right].
$$
where $\varphi_0=\arg(G(k_0))$. Then we have
$$
A(k_0,k_0,l,\lambda)=\lambda|G(k_0)|e^{2\pi i\Phi(k_0,l,\lambda)}
$$
and
$$
\log(A_0) + 2\pi i j = \log(\lambda |G(k_0)|) +2\pi i(j+\Phi(k_0,l,\lambda))
$$
so that
\begin{align*}
\Re(k_{j,l}) &\sim k_0 
+ \frac{2\pi}{\lambda}(j+\Phi(k_0,l,\lambda)) 
-\frac{1}{\lambda} \arctan\left(\frac{2\pi (j+\Phi(k_0,l,\lambda))}{\log(\lambda |G(k_0)|)}\right)\\
\end{align*}
and
\begin{align*}
\Im(k_{j,l}) &\sim
-\frac{\log(\lambda |G(k_0)|)}{\lambda}
+\frac{1}{2\lambda}\log\left( \log(\lambda |G(k_0)|)^2 + 4\pi^2(j+\Phi(k_0,l,\lambda))^2 \right)
\end{align*}
So we get that the real parts have a spacing of $\sim 2\pi/\lambda=p\pi/L$, i.e., $p$ times wider than in case 1. But there are $p$ sequences corresponding to $l=0,\ldots,p-1$ so the counting remains the same. The imaginary parts are most negative when $j=0$, which corresponds to real parts close to $k_0$. 

\bigskip
\setlength{\fboxsep}{10pt}
\fbox{\begin{minipage}{40em}
Here the approximate resonances (diamonds) are compared to the first two terms in their expansions (crosses).

\vskip -2.5cm\

\begin{center}
\includegraphics[width=9cm]{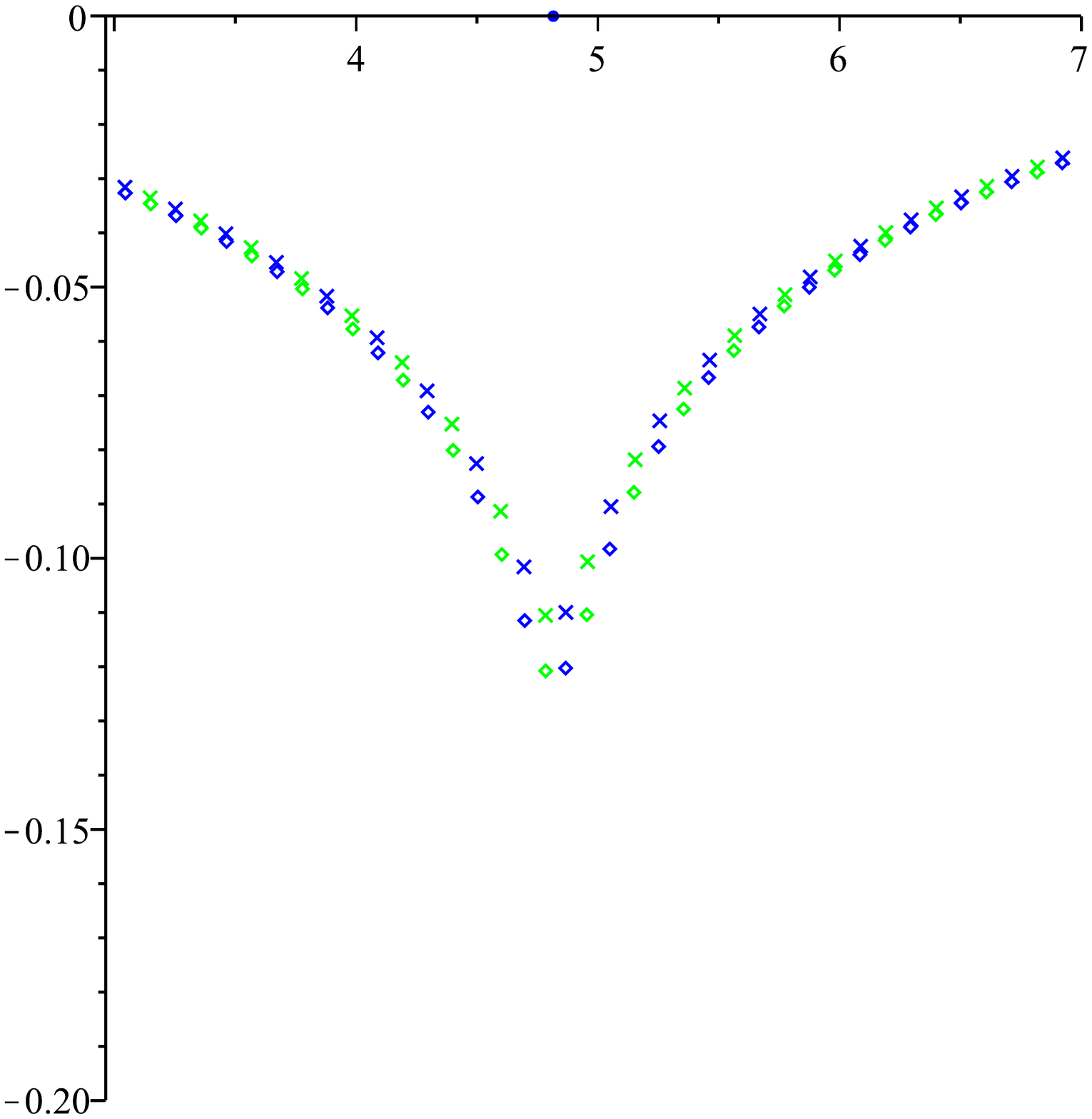}
\end{center}
\end{minipage}}
\bigskip\bigskip

\noindent{\bf Case 3: $\Im k_0 < 0$ and $k_0$ is a pole of $f$}.
\bigskip

As in Case 2, {a point $k$ near $k_0$ } is a resonance if and only if it satisfies 
$$
k=k_0 + \frac{1}{i\lambda}W(A(k,k_0,l,\lambda))
$$
for some branch $W$ of the Lambert $W$ function, with
$$
A(k,k_0,l,\lambda) = i\lambda e^{-i\lambda k_0}\omega^lG(k),
$$
where as before, $G(k)$ is a branch of $g(k)^{1/p}$ and $\omega=e^{2\pi i/p}$.
We may define approximate resonances as before as
$$
k_{j,l} = k_0 + \frac{1}{i\lambda}W(A(k_0,k_0,l,\lambda)).
$$
The difference is that now $|A(k,k_0,l,\lambda)|=\lambda e^{-|\Im k_0|\lambda}|G(k)|$ is exponentially small in $\lambda$. This means that we must consider the behaviour of branches $W_j(z)$ of the Lambert W function near $z=0$. The principal branch $W_0(z)$ is analytic in a disk of radius $1/e$ about $z=0$ with leading behaviour $W_0(z) = z + O(z^2)$.
We will use that there is a constant $C$ such that  for $|z|<1/10$,
\begin{equation}\label{w0taylor}
|W_0(z)-z| \le C |z|^2.
\end{equation}

The other branches $W_j(z)$ for $j\ne 0$ all behave like $\log(z)$ and are given by the convergent expansions above. The differing behaviours result in only the $p$ approximate resonances with $j=0$ being associated to resonances near $k_0$.
Let ${\cal C}(k_0,\lambda) = \{k_{0,l}, l=0\ldots p-1\}$ denote these  approximate resonances.

\begin{proposition}
Suppose $f(k)$ has a pole of order $p$ at $k_0\in\bbC_-$.  Let $\epsilon \in(0,1/2)$. Then there exists $\delta>0$ and $\lambda_0$ such that for all  $\lambda>\lambda_0$
\begin{enumerate}[label=(\roman*)]
\item If $k_{0,l}\in B(k_0,\delta)\cap {\cal C}(k_0,\lambda)$ then there is a $k \in {\cal R}(L)$ with $|k-k_{0,l}|<\epsilon e^{-|\Im(k_0)|\lambda}$.
\item If $k\in B(k_0,\delta)\cap {\cal R}(L)$ then there is a $k_{0,l}\in {\cal C}(k_0,\lambda)$  with $|k-k_{0,l}|<\epsilon e^{-|\Im(k_0)|\lambda}$.
\end{enumerate}
\end{proposition}
\begin{proof}
Choose $\delta$ small enough to ensure that $G(k)$ is analytic and $|G(k)-G(k_0)|\le \epsilon/2$ for   $k\in B(k_0,2\delta)$ and $\delta < |\text {Im} k_0|$. Choose $\lambda_0$ sufficiently large so that $\lambda>\lambda_0$ implies
$$
\lambda e^{-|\Im k_0|\lambda}(|G(k_0)|+1)<\frac{1}{10},\quad
e^{-|\Im k_0|\lambda}<\delta,\quad
\lambda e^{-|\Im k_0|\lambda}2C(|G(k_0)|+1)^2<\epsilon/2,
$$
where $C$ is the constant in \eqref{w0taylor}.

Then for  $k\in B(k_0,2\delta)$ and  $\lambda>\lambda_0$,
$$
|A(k,k_0,l,\lambda)|=\lambda e^{-|\Im k_0|\lambda}|G(k)|\le \lambda e^{-|\Im k_0|\lambda}(|G(k_0)|+1)<\frac{1}{10}
$$
This implies that $A(k,k_0,l,\lambda)$ lies in the region of analyticity for $W_0$ and that $|W_0(A)-A| \le C |A|^2$ for both $A=A(k,k_0,l,\lambda)$ and $ A=A(k_0,k_0,l,\lambda)$.

Now assume that we are given $k_{0,l}\in{\cal C}(k_0,\lambda)\cap B(k_0,\delta)$. We will apply Rouch\'e's theorem to 
$$\phi(k) = k - k_0 - \frac{1}{i\lambda}W_0(A(k,k_0,l,\lambda))$$
and 
$$\xi(k) =k-k_{0,l} = k - k_0 - \frac{1}{i\lambda}W_0(A(k_0,k_0,l,\lambda)).$$
on $B(k_{0,l},\epsilon e^{-|\Im(k_0)|\lambda})$.
Clearly $\xi(k)$ is analytic on this ball and
$|\xi(k)| =\epsilon e^{-|\Im(k_0)|\lambda}$ on the boundary. On the other hand, given that $k_{0,l}\in B(k_0,\delta)$ we find that  points $k\in B(k_{0,l},\epsilon e^{-|\Im(k_0)|\lambda})$ obey
$$
|k-k_0| \le |k-k_{0,l}|  + |k_{0,l}-k_0|\le  \epsilon e^{-|\Im(k_0)|\lambda} + \delta <2\delta
$$
provided $\lambda>\lambda_0$. 
So $k\in B(k_0,2\delta)$. This implies that $\phi(k)$ is analytic in neighbourhood of $B(k_{0,l},\epsilon e^{-|\Im(k_0)|\lambda})$.

Let $A(k)$ denote $A(k,k_0,l,\lambda)$. For $k\in B(k_{0,l},\epsilon e^{-|\Im(k_0)|\lambda})$ and $\lambda>\lambda_0$ we have
\begin{align*}
|\phi(k)-\xi(k)| &= \frac{1}{\lambda} |W_0(A(k))-W_0(A(k_0))|\\
&\le\frac{1}{\lambda}\left( |A(k)-A(k_0)| + C\left(|A(k)|^2+|A(k_0)|^2\right)\right)\\
& \le \frac{1}{\lambda} \Big(\lambda e^{-|\Im(k_0)|\lambda}|G(k)-G(k_0)|+ 2C(|G(k_0)|+1)^2\lambda^2 e^{-2|\Im(k_0)|\lambda}\Big)\\
&\le \left( \frac{\epsilon}{2} +2C(|G(k_0)|+1)^2 \lambda e^{-|\Im(k_0)|\lambda}|\right)e^{-|\Im(k_0)|\lambda}\\
&<\epsilon e^{-|\Im(k_0)|\lambda}\\
\end{align*}
Now (i) follows from Rouch\'e's theorem.

Now suppose that $k$ is a resonance. Then
$
k=k_0 + \frac{1}{i\lambda}W_j(A(k,k_0,l,\lambda))
$
for some $j,l$. When $j=0$ our previous estimate shows 
$$
|k-k_{0,j}| \le  \frac{1}{\lambda} |W_0(A(k))-W_0(A(k_0))|<\epsilon e^{-|\Im(k_0)|\lambda}
$$
provided $k\in B(k_0,\delta)$ and $\lambda>\lambda_0$. So (ii) holds in this case.

When $j\ne 0$ the values of $W_j$ are bounded away from zero and we have
$$
e^{\Re(W_j(z))}=\frac{|z|}{|W_j(z)|} \le C |z|
$$
Since $|A(k,k_0,l,\lambda)| < C(k_0) \lambda e^{-|\Im k_0|\lambda}|$ we conclude
$$
\Re(W_j(A(k,k_0,l,\lambda)))<-|\Im k_0|\lambda+\ln(C(k_0)\lambda).
$$
This implies that
$$
|k-k_0|\ge\frac{1}{\lambda}|\Re(W_j(A(k,k_0,l,\lambda)))|\ge |\Im k_0|
-\ln(C(k_0)\lambda)/\lambda  > \delta
$$
for $\lambda$ large enough.  Thus (ii) holds vacuously.

\end{proof}

\begin{proof} \textbf{of Theorem \ref{thm:counting}}
The counting part of the proof follows from the formulas for the approximate resonances in Case 1) and Case 2) and from Propositions \ref{prop:nopole} and \ref{asympt1}.  Much of what we show below follows from the estimates above but we give a simple self-contained proof.  Suppose $f$ is analytic at the point $k_0 \in \mathbb{R}$.  Then if $\text{Im} k < -c_1/L$ and $k$ is a resonance near $k_0$ we have 
$$|f(k)| = e^{-2\text{Im} k L} > e^{2c_1}$$  which is false for $c_1$ large enough and $k$ in a fixed $L$ - independent neighborhood of $k_0$.  If $k_0$ is a pole of $f$ of order $p$ then $|f(k)| \le  c|k-k_0|^{-p}$ in a neighborhood of $k_0$.  If $k$ is a resonance in this neighborhood with $\text{Im} k < - (p/2)L^{-1} \log L$ then 
$$ |f(k)| = e^{-2\text{Im} k L} > e^{p\log L} = L^p .$$  But $$ |f(k)| \le c/|k-k_0|^p < c(2p^{-1})^p (L/\log L)^p,$$ a contradiction for large $L$.
\end{proof}

\section{Resonances as $L\rightarrow\infty$ when $V_{2,L}(x) = e^{-cL}\delta(x-L)$}\label{mudelta}

In this section we determine the asymptotic positions of the resonances of $H_L=-d^2/dx^2 + V_1(x) + e^{-cL}\delta(x-L)$for large $L$. The analysis is similar to the previous section, so we will omit many details. {At the end of this section we make some comments about how things change when instead of $ e^{-cL}\delta(x-L)$ we use $\mu(L) V_2(x-L)$ where $V_2$ }{ satisfies the same hypotheses as before} {and $\mu(L)$ can be $e^{-cL}$ or decrease faster or slower than an exponential.  These comments are based on calculations presented in an Appendix.}

{For now we consider the moving delta function.}  The functions $f_1(k)$ and $f_2(k)$ are defined by \eqref{eqn:f} as before. But now $f_2$ depends on $L$. We can compute explicitly that
$$
f_2(k) = \frac{e^{-cL} - 2ik}{e^{-cL}}.
$$
The zero of $f_2(k)$ gives the position of the single resonance of $-d^2/dx^2 + e^{-cL}\delta(x)$ at $k=-ie^{-cL}/2$. There are no reflectionless points for $V_2=\delta$ and therefore no poles for $f_2$. Equation \eqref{eqn:R} can now be written
\begin{equation}\label{eqn:R2}
e^{(2ik-c)L} = f_1(k) (e^{-cL} - 2ik)
\end{equation}

We find resonance free regions as before. Let
$$
U_1(a,A) = \{k\in\bbC_- : \Im(k) > -c/2 + a, |f_1(k)| > 1/A \hbox{\ and\ }|k| > 1/A\}
$$
for $a,A>0$. Now we are excluding neighbourhoods of resonances of $H$, since these occur at the zeros of $f_1$.
Then we let 
$$
U_2(a,A) = \{k\in\bbC_- : \Im(k) < -c/2 - a, |f_1(k)| < A \hbox{\ and\ }|k| < A\},
$$
again for $a,A>0$. Here we are excluding neighbourhoods of reflectionless points of $H$ as before.
\begin{proposition}\label{leaving2}
There exists $L_0$ such that $U_1(a,A)\cap{\cal R}(L) = U_2(a,A)\cap{\cal R}(L) =\emptyset$ when $L>L_0$.
\end{proposition}
\begin{proof}
If $k\in U_1(a,A)$ is a resonance then using \eqref{eqn:R2} we find 
$$
e^{-2aL}\ge |e^{(2ik-c)L}| = |f_1(k) (e^{-cL} - 2ik)|\ge \frac{1}{A}\left(\frac{2}{A}-e^{-cL}\right)
$$
If $k\in U_2(a,A)$ is a resonance then
$$
e^{2aL} \le |e^{(2ik-c)L}| = |f_1(k) (e^{-cL} - 2ik)|\le A(1+2A)
$$
both these inequalities are impossible for $L>L_0$ for some $L_0(a,A,c)$.
\end{proof}

{\begin{proposition}\label{converging}
If $k$ is a resonance of $H$ with $ \text{Im} (k) > -c/2$, then $k$ is a limit of resonances of $H_L$ as $L \to \infty$. If $k$ is a reflectionless point of $H$ with $ \text{Im} (k) < -c/2$ then $k$ is a limit of resonances of $H_L$ as $L \to \infty$.  The resonances of $H$ with $\text{Im} (k) <-c/2$  are not limits of resonances of $H_L$.  The resonances of $H_L$ accumulate densely on the line $\text{Im} (k) = -c/2$.
\end{proposition}}

{We do not explicitly give the proof of Proposition \ref {converging} but it will be clear as we treat the six cases below. }

Retracing our steps {from Section \ref{no coupling constant}}, we now find all the resonances near some point $k_0$ in the closed lower half plane. In view of \ref{leaving2} we consider six cases: $k_0 = 0$, $\Im(k_0)>-c/2$ and $f_1(k_0)=0$, $\Im(k_0)=-c/2$ and $0<|f_1(k_0)|<\infty$, $\Im(k_0)=-c/2$ and $f_1(k_0)=0$, $\Im(k_0)=-c/2$ and $k_0$ is a pole of $f_1$ and $\Im(k_0)<-c/2$ and $k_0$ is a pole of $f_1$. In each case  {(except for $k_0 = 0$ where we will do more)} we will identify the approximate resonances in terms of the log or the Lambert $W$ function but omit the error estimates, which can be carried out as in the previous section. The error estimates say that near $k_0$ every resonance has a nearby approximate resonance and vice versa.

\bigskip
\noindent{\bf Case 1: $k_0 = 0$}\\
\medskip

We know that $|f_1(0)| \ge 1$ and that generically $f_1(0) =1$.  But in fact there are situations where $f_1$ has a pole at $0$.  So write equation (\ref{eqn:R2}) as 
$$e^{2ikL} = \frac{g(k)}{k^p}(1-2ike^{cL})$$ where $g$ is analytic in a neighborhood of $|k| \le \delta < c/2$,  $p$ is a non-negative integer and $|g(k)| \ge c_0 > 0$ for $|k|\le \delta$.  Choose $\epsilon \in (2\delta, c)$ and let $|k|\le\delta$ . Then if $|k| > e^{-(c-\epsilon)L}$ the left side of (\ref{eqn:R2})
 has modulus $\le e^ {2\delta L}$ while the modulus of the right side is $\ge \frac{c_0}{\delta^p} (2e^{\epsilon L} - 1)$ which is false for large $L$ if $k$ satisfies  (\ref{eqn:R2}).  Thus for large $L$ any solution to  (\ref{eqn:R2})  with $|k| \le \delta$ satisfies $|k| \le e^{-(c-\epsilon)L}$.  Iterating this argument once more with an improved estimate on the modulus of the left side of  (\ref{eqn:R2}) shows that for large $L$ any solution to  (\ref{eqn:R2})  with $|k| \le \delta$ satisfies $|k| \le R e^{-cL}$ with $R = 1/2 + c_0^{-1}$. 
Let $\xi = ke^{cL}$ and $$f(\xi) = (2i)^{-1}\big(1 - e^{2ikL}(\xi e^{-cL})^p (g(\xi e^{-cL})^{-1} \big ).$$  Then (\ref{eqn:R2}) becomes $f(\xi) = \xi$.  $R$ has been chosen so that it is clear that $f$ maps the closed ball $\overline{B(0,R)}$ into itself. In this closed  ball we can see that $|f'(\xi)| \le C_1 Le^{-cL}$ so that for large $L$ there is a unique solution to  (\ref{eqn:R2}) for $|k| \le \delta$.  If $f_1(0) = 1$ then this is the $0$ solution.  Thus generically $k=0$ is the only solution to  (\ref{eqn:R2})  for $|k| \le \delta$. If $0 < |f_1(0) -1| <\infty$, let  $\xi_1 = (2i)^{-1}(1-f_1(0)^{-1})$ and compute that $\xi_2 = f(\xi_1) = \xi_1 + O(Le^{-cL})$.  Then if $\xi^* =f(\xi^*)$ we have $|\xi^* -  \xi_2| = |f(\xi^*) - f(\xi_1)| \le \max_{t\in [01]} |f'( (1-t)\xi^* + t \xi_1)||\xi^* - \xi_1|  = O(Le^{-cL})$.  Thus if $0< |1-f_1(0)| < \infty$ the unique solution to  (\ref{eqn:R2}) with $|k| \le \delta$ is $k = (2i)^{-1}(1-f_1(0)^{-1})e^{-cL} + O(Le^{-2cL})$.  Finally suppose $p\ge 1$.  Then $\xi^* = f(\xi^*) = 1/2i + O(e^{-pcL }) $ so the unique solution  to (\ref{eqn:R2}) with $|k| \le \delta$ is $k = (2i)^{-1}e^{-cL} + O( e^{-(p+1)cL})$.\\

To summarize: In the generic case where $f_1(0) = 1$ the unique solution with $|k| < \delta$ is $k=0$. \\

If $|1-f_1(0)| <\infty$ then the unique solution with  $|k| < \delta$ satisfies

$$ k = (2i)^{-1}(1-f_1(0)^{-1})e^{-cL} + O(Le^{-2cL})$$

and if $f_1$ has a pole of order $p$ at $0$ the unique solution with $|k| < \delta$ satisfies

$$k = (2i)^{-1}e^{-cL} + O( e^{-(p+1)cL}).$$

The case where $k_0=0$ and $f_1$ has a pole at $k_0$ is different from  {all} the other cases with poles  {in Section \ref {no coupling constant} and below},  because there is only one resonance near $k_0$ even when  $p > 1$.


\bigskip
\noindent{\bf Case 2: $\Im(k_0)>-c/2$ and $f_1(k_0)=0$}\\
\medskip

These are the resonances of $V_1$ above the line $\Im(k)=-c/2$. Write $f_1(k) = (k-k_0)^p g(k)$ where $g(k)$ is analytic at $k_0$ and $g(k_0)\ne 0$. Then \eqref{eqn:R2} becomes
$$
(k-k_0)^p e^{(-2ik+c)L} = \frac{1}{g(k)(e^{-cL}-2ik)}.
$$
Let $\lambda =2 L/p$ and $\omega = e^{2\pi i/p}$. Let $G(k)$ be a fixed branch of $(g(k)(e^{-cL}-2ik))^{1/p}$ analytic near $k_0$. Then the equation above is equivalent to the $p$ equations
$$
(k-k_0) e^{(-ik+c/2)\lambda} = \frac{\omega^l}{G(k)},
$$
for $l=0\ldots p-1$, which we rewrite as
$$
-i\lambda(k-k_0) e^{-i(k-k_0)\lambda} = \frac{-i\lambda\omega^le^{(ik_0-c/2)\lambda}}{G(k)}.
$$
Solutions to these equations are the same as solutions to
$$
k = k_0 - \frac{1}{i\lambda}W_j\left(\frac{-i\lambda\omega^le^{(ik_0-c/2)\lambda}}{G(k)} \right)
$$
for $l=0\ldots p-1$ and $j\in\bbZ$. Since the argument of $W_j$ is exponentially small, we are in the situation where only $j=0$ results in an approximate resonance. So we have $p$ approximate resonances
$$
k_{0,l} =  k_0 - \frac{1}{i\lambda}W_0\left(\frac{-i\lambda\omega^le^{(ik_0-c/2)\lambda}}{G(k_0)} \right)
$$
close to $k_0$. 

\bigskip
\noindent{\bf Case 3: $\Im(k_0)=-c/2$ and $0<|f_1(k_0)|<\infty$}\\
In this case we can write \eqref{eqn:R2} as
$$
e^{2i(k-k_0)L} = e^{(c-2ik_0)L}f_1(k)(e^{-cL}-2ik) = e^{-2i\Re(k_0)L}f_1(k)(e^{-cL}-2ik)
$$
and take the $\log$. This results in approximate resonances at
$$
k_j = k_0 + \frac{1}{2iL}\log(e^{-2i\Re(k_0)L}f_1(k_0)(e^{-cL}-2ik_0) + \frac{j\pi}{L}.
$$

\bigskip
\noindent{\bf Case 4: $\Im(k_0)=-c/2$ and $f_1(k_0)=0$}\\
\medskip

This is like Case 2 except that $i(k_0-c/2)=i\Re(k_0)$ is purely imaginary so we no longer have exponential decay in $\lambda$ in the argument of the Lambert $W$ function. Instead we have linear growth which means that all the $W_j$'s will contribute approximate resonances near $k_0$.  Thus, the approximate resonances are
$$
k_{j,l} =  k_0 - \frac{1}{i\lambda}W_j\left(\frac{-i\lambda\omega^le^{i\Re(k_0)\lambda}}{G(k_0)} \right)
$$

\bigskip
\noindent{\bf Case 5: $\Im(k_0)=-c/2$ and $k_0$ is a pole of $f_1$}\\
\medskip

Cases 5 and 6 are related analogously to Cases 2 and 4. When $\Im(k_0)=-c/2$ we will obtain approximate resonances
$$
k_{j,l} =  k_0 + \frac{1}{i\lambda}W_j\left(i\lambda\omega^le^{-i\Re(k_0)\lambda}G(k_0) \right)
$$
where $f_1(k)=(k-k_0)^{-p}g(k)$ and $G(k)$ and $\omega$ are defined as before. 

\bigskip
\noindent{\bf Case 6: $\Im(k_0)<-c/2$ and $k_0$ is a pole of $f_1$}\\
\medskip

These are the reflectionless points for $V_1$  below the line $\Im(k)=-c/2$. Close to these points we will get $p$ approximate resonances
$$
k_{0,l} =  k_0 + \frac{1}{i\lambda}W_0\left(i\lambda\omega^le^{(c/2-ik_0)\lambda}G(k_0) \right)
$$

{We now replace $e^{-cL}\delta(x-L)$ with $\mu V_2(x-L)$.  We will take $\mu = \mu(L)$ with $\lim_{L \to \infty} \mu(L) = 0$.  We take  $V_2$  as before  with compact support in $(0,x_1)$.  From the Appendix in Section 9, The function $f_2$ for small coupling,  we find}{
\begin{equation}\label{eqn:muV2}
\mu f_2(k,\mu) =  -2ik/I(k) + O(\mu).
\end{equation}}
where $I(k) = \int _{\mathbb R} V_2(x) e^{2ikx} dx$.  The $O(\mu)$ term is uniform on compact subsets of $k$ which do not contain $k=0$.

{Here we have indicated explicitly that $f_2$ depends on $\mu$ as well as $k$.  This expression should be compared to the expression for $f_2$ obtained above when  $\mu V_2(x-L)$  is replaced by $e^{-cL}\delta(x-L)$, namely
$$e^{-cL}f_2(k) = -2ik + e^{-cL}$$}
{With $\mu V_2$ instead of $e^{-cL}\delta$, (\ref{eqn:R2}) becomes
$$\mu e^{2ikL} = f_1(k)\mu f_2(k,\mu)$$}

{It is clear from (\ref{eqn:muV2}) that if $\mu(L) \to 0 $ faster than any exponential, then for any compact set  $K \subset \mathbb C \setminus\{0\}$, all resonances of $H_L = H + \mu(L) V_2(x-L)$  either leave $K$ or converge to resonances of $H$ in $K$.  In fact if $k_0$ is a resonance of $H$ (thus a zero of $f_1$), there is a resonance of $H_L$ converging to $k_0$.}

{Consider now a compact subset $K$ of $ \{k \ne 0: I(k) \ne 0, \text{Im} k  < -a < 0\} $.  (We assume $ I(k)  \ne 0$ and $k \ne 0$ for simplicity.)  Then if $\mu(L) \to 0$  slower than any exponential, the resonances of $H_L$ either leave $K$ or converge to a reflectionless point of $H$.  Resonances of $H_L$ converge to the real axis in a similar way as is discussed in the case that $\mu(L) = 1.$  (For example if $f_1(k)$  is analytic at $k_0 \in \mathbb R$ the resonances near $k_0$ are uniformly displaced by approximately $-(2iL)^{-1}\log \mu(L)$ but the separation between them is still approximately $\pi/L$. Again we assume $I(k_0) \ne 0$ and $k_0 \ne 0$ .)}\\

{If $\mu(L) = e^{-cL}$, the analysis is very similar to that given earlier in this Section but instead of equation (\ref{eqn:R2}) we have
$$e^{(2ik - c)L} = [f_1(k) (-2ik)/I(k)](1 + O(e^{-cL}).$$ 
We have assumed we are in a neighborhood of a point $k_0 \ne 0$ where $I(k_0) \ne 0$. The basic phenomenology is the same as with $e^{-cL}\delta(x-L)$.  We omit the proofs.}

\section {Exponential decay - lost resonances}

In this section we consider the Hamiltonian $H_L = H + V_{2,L}(x)$ where $H = -d^2/dx^2 + V_1(x)$ and  $V_{2,L}(x) = V_2(x-L)$.  We assume $V_1$ and $V_2$ are real  and that at least $V_2$ is bounded and that $\lim_{|x| \to \infty} V_2(x) = 0$.  We are interested in a situation where $\psi$ is a long lived (normalized) state for $H$ in the sense that $(\psi, e^{-itH}\psi) \sim e^{-itE} $ for a long time.  Here $E$ is a resonance for $H$. We know that the resonances of $H$ disappear when we perturb $H$ by adding $V_{2,L}$ even when $L \to \infty$.  Nevertheless we expect  $(\psi, e^{-itH_L}\psi) \sim e^{-itE} $ for a long time if $L$ is large.

The proof of this boils down to showing $||e^{-itH_L}\psi - e^{-itH}\psi||$ is small for a long time $T$ as long as $L$ is large enough.  We give two results of this nature.  The first relies on a compactness argument and is not very sensitive to the geometry of the problem and thus does not find any relation between $T$ and $L$.  However it works for any $\psi \in L^2(\mathbb R)$ and no further assumptions on the potentials.  The second result is stronger and harder to prove. It gives an explicit upper bound for  $T$ in terms of $L$ which is needed to get a good estimate of the error.  But it assumes more about $\psi$ and the potentials.  

\begin{theorem}\label{feelingV_L-1}
Suppose $V_j$, $j=1,2$ are real and measurable and $H = -d^2/dx^2 + V_1$ is self-adjoint.  Suppose $V_2$ is bounded and $\lim_{|x| \to \infty} V_2(x) = 0$.  Fix $\psi \in L^2(\mathbb R)$.  Given $\epsilon, T >0$ there exists $L_0$ such that 

$$||e^{-itH_L}\psi - e^{-itH}\psi|| < \epsilon$$
if $|t| \le T$ and $L\ge L_0$.

\end{theorem}

\begin{proof}
We have 
$$e^{-itH_L} \psi- e^{-itH} \psi = -i\int_0^t e^{-i(t-s)H_L} V_{2,L}e^{-isH}\psi ds $$
so that 
\begin{equation}\label{basic equation}
||e^{-itH_L} \psi - e^{-itH} \psi|| \le |\int_0^t ||V_{2,L}e^{-isH}\psi || ds|
\end{equation}
Since $s-\lim_{L \to \infty} V_{2,L} = 0$ and $\{e^{-isH}\psi : |s| \le T\}$ is compact, $\lim _{L\to \infty} ||V_{2,L}e^{-isH}\psi|| = 0$ uniformly for $|s|\le T$.  Thus we can find $L_0$ such that 
$$||V_{2,L}e^{-isH}\psi|| < \epsilon/T$$ for $|s| \le T$ which gives the result.
\end{proof}

\begin{theorem} \label{feelingV_L-2}
Suppose $V_j, j=1,2$ are real functions in $C_0^{\infty}(\mathbb R)$.
Suppose $\psi \in \mathcal {S}(\mathbb R)$.  Then given $\delta$ with $0< \delta <<1$ there is a constant $c = c_{\delta} > 0$ such that

$$||e^{-itH_L} \psi - e^{-itH} \psi|| = O(L^{-n})$$
for all $n$ if $0 \le t \le cL^{1-\delta}$.  

\end{theorem}

\begin{remark} 

\begin{enumerate}

  \item
 We have in mind that $H$ might have a resonance close to the real axis and that there will be normalized states  $ \psi \in \mathcal {S}(\mathbb R)$ satisfying 

$$ |(\psi, e^{-itH}\psi) - e^{-itE}|  < \epsilon$$
for some small $\epsilon$ and all $t \ge 0$. Here $E$ is a resonance for $H$ with small negative imaginary part.  The bound is only useful for  $|e^{-iEt}| > \epsilon$ or $0\le t \le T < (\log \epsilon^{-1})/|\text{Im} E|$.  According to Theorems \ref{feelingV_L-1} and \ref{feelingV_L-2} if $t$ is in this range  the same bound will hold  
if $H$ is replaced by $H_L$ as long as  $L$ is large enough.

\item
In Theorem \ref{feelingV_L-2}  we assumed $\psi \in  \mathcal {S}(\mathbb R)$ and that the $V_j$ are smooth with compact support in order to get a simple result.  With suitable weaker assumptions there will be similar but weaker bounds.

\item

If $\psi$ has bounded energy, $H$, and $L$ is large it takes some time before $e^{-itH_L} \psi$ will feel $V_{2, L}$ so that we would expect that the bound in Theorem \ref{feelingV_L-2} would be valid for $0 \le t \le cL$ for some constant $c$ depending on the maximum velocity (= twice the momentum)  in $\psi$.  That this is true will be clear from the proof.
\end{enumerate}

\end{remark}

The proof of Theorem \ref{feelingV_L-2} is given in  the  first Appendix.

\section{Resonances and long lived states}

What is the physical meaning of the resonances that are crowding together and approaching a horizonal line? It is not completely clear to us, but we offer the following discussion.

If $k_0$ is a resonance with resonant energy $k_0^2=\lambda_0-i\delta_0$, $\delta_0 > 0$, and $\delta_0$ is small, then $k_0$ should correspond to a long lived resonant state. What is meant by this is open to discussion. A possible definition is a normalized state $\varphi$ satisfying
\begin{equation}\label{resstate}
\sup_{|t|\le T}\left |\left\langle \varphi, e^{-itH}\varphi\right\rangle - e^{-it\lambda-|t|\delta}\right| \le \epsilon
\end{equation}
with possible conditions on $T$, $\epsilon$ and the support (or variance) of $\varphi$ { in $x$ - space}. 

Lavine [L] showed that if $k_0=\sigma_0-i\epsilon_0$ is a resonance and $\varphi$ is the (exponentially growing) outgoing solution of \eqref{eqn:SSE}, cut off so its support matches that of the potential, and normalized, then \eqref{resstate} holds with $T=\infty$ and with $\epsilon=O(\epsilon_0/\sigma_0)\log(\sigma_0/\epsilon_0)$ for small $\epsilon_0/\sigma_0$ with explicit constants. Lavine proved the result in the context of half line scattering. A version for the whole line that applies the Hamiltonians we consider is given in Appendix 2. 

Lavine's result gives us a resonant state for each of the resonances in our examples that are crowding  {onto} the real axis. But since the supports of the potentials in our examples have width $L$, the resulting resonant states, which have the same support, are not well localized. Non-localized states satisfying \eqref{resstate} can exist even when $H$ has potential $V=0$ and no resonances at all, as was pointed out by Lavine [L]. We need only find $\varphi$ such that the spectral measure $d\mu_\varphi$ is close to the Lorenzian $d\mu_L$ corresponding to $\lambda_0-i\delta_0=k_0^2$. We can't do this exactly because the spectral measure  $d\mu_\varphi$ is supported in $[0,\infty)$ and $d\mu_L$ has non-zero density everywhere on $\bbR$. But by making $\lambda_0$ large, we can make $\epsilon$ as small as we like. But the resulting $\varphi$ is very spread out. {Thus in search of a physical meaning for these resonances which are densely crowding on the real axis one might try to superpose them to obtain states which shuttle back and forth between the supports of $V_1$ and $V_{2,L}$ a few times before eventually the whole wave function is transmitted through the potentials.}

Does an estimate like \eqref{resstate} ever signal the existence of a resonance or eigenvalue?
If $\epsilon=0$ and $T=\infty$ then the answer is yes. In this case the spectral measure $d\mu_\varphi$ satisfies $\int_\bbR e^{its}d\mu_\varphi(s) = \left\langle \varphi, e^{-itH}\varphi\right\rangle = e^{-it\lambda_0-|t|\delta_0}$ which implies that $d\mu_\varphi = d\mu_L$, where $d\mu_L$ is the Lorenzian with parameters $\lambda_0$ and $\delta_0$.
Since the spectral measure is bounded below, this is impossible unless $\delta_0=0$. In this limiting case  $d\mu_L$
is a delta function supported on $\{\lambda_0\}$. Thus $\varphi$ is an eigenfunction of $H$ with eigenvalue $\lambda_0$. 

If $\epsilon>0$, then an estimate like \eqref{resstate} even with a localized $\varphi$ need not signal the existence of resonances near $k_0$. Consider $H=-d^2/dx^2 + V_1$ for a potential $V_1$ supported in $[-1,1]$. Lavine's theorem gives us a resonant state $\varphi$ also supported in $[-1,1]$ satisfying \eqref{resstate} with $T=\infty$ and $\epsilon$ depending only on $k_0$. The estimate \eqref{resstate} will continue to hold, with slightly adjusted constants if we mollify $\varphi$ to put it in $\cal S$. But then, Theorem \ref{feelingV_L-2} implies that \eqref{resstate} also continues to hold when $H$ is replaced with $H_L=H+V_2(x-L)$, although we must change $T$ from $\infty$ to $cL^{1-\delta}$. As $L$ increases this resonant state lives for an increasingly long time without becoming delocalized. At the same time, the resonances of $H_L$ are moving away from $k_0$.

\section{Appendix: Proof of Theorem \ref{feelingV_L-2}}

Our proof of Theorem \ref{feelingV_L-2} uses the basic equation (\ref{basic equation}).  Thus we need to bound  $||V_{2,L}e^{-isH}\psi||$.

For this we need a propagation estimate to show that $e^{-isH} \psi(x)$ remains small for $x \ge L$ for times $s$ for which the velocities in $\psi$ do not allow propagation of $\psi $ from around the origin where it is localized initially to  $x \ge L$.  We first make an energy cut-off for $\psi$ so that we have an effective cut-off in the velocity.  We finally estimate the remainder using the fact that $\psi$ is also well localized in energy. 

Let $\tilde {h}_j, j = 1,2$ be two real non-negative $C^{\infty}(\mathbb R)$ functions such that $\tilde{h}^2_1 + \tilde{h}^2_2 = 1$ and $\tilde{h}_1(t) = 1$ for $-\infty < t < 1$ and  $\tilde{h}_1(t) = 0$ for $ 1+\epsilon < t < \infty$ with $\epsilon > 0$.  Let $k_j(t) = \tilde {h}_j(t/K)$ where we will later choose $K$ to increase with $L$.  Let $\psi_K = k_1(H)\psi$ and $\psi_{K,t} = e^{-itH}\psi_K$. Let $h_j(t) = \tilde {h}_j(t/(1+\epsilon)K)$

Let $v = 2(1+\epsilon)\sqrt K$.  ($2\sqrt {(1+\epsilon)K}$ is the maximum possible velocity in $\psi_K$  if we ignore $V_1$).  Let $\langle x \rangle = \sqrt{|x|^2 + 1}$.

\begin{proposition}

For any $\alpha > 0$ and $t\ge 0$

\begin{equation} \label{propest}
||1_{\{|x| > v(1+ 2\epsilon)(t+1)\ge 0 \} }\langle x \rangle^{\alpha/2} \psi_{K,t}|| \le (1+\epsilon^{-1})^{\alpha}||\langle x \rangle^{\alpha/2} \psi_{K}|| + C_{\alpha}(K)
\end{equation}
where  $\lim_ {K \to \infty} C_{\alpha}(K) = 0$.
\end{proposition}

An estimate of this form appears in \cite {S} where Skibsted proves several propagation estimates for N-particle problems.  The above estimate is essentially in \cite {S} but for an $h_1$ which vanishes in a neighborhood of the thresholds of the problem ($0$ in our case).  This restriction was necessary for other propagation estimates but not for this one (as Skibsted knew \cite { S1}).  Control of the constants which occur in the proof of the result is essential for us and this needs a bit more care than is present in the proof of  \cite {S} .  For these reasons and for completeness we present a proof of the proposition (following the method in \cite {S}).  We do not assume that we are in one dimension.

\begin{proof}

We first introduce the propagation observable:  $A(\tau) = \tau v - \langle x \rangle $, where $\tau = t + 1$.  For each $\alpha \ge 0$ we will need a function $$g(u, \tau) = -(-u)^{\alpha} \chi(u/\tau)$$ where $\chi$ satisfies the following:  $u\chi'(u) + \alpha\chi(u)= \tilde \chi^2(u)$ with  $\chi' \le 0$, $\tilde\chi \ge 0$, and $\tilde \chi \in C^{\infty}(\mathbb R)$, $\chi(u) = 1, u \le -2\epsilon v,  \chi(u) = 0, u \ge -\epsilon v$ for $\epsilon > 0$.  Such a $\chi$ can be shown to exist for any $\alpha \ge 0$ and $\epsilon v > 0$.
We calculate

$$\partial g(u,\tau)/ \partial u = (-u)^{\alpha - 1}[\alpha \chi(u/\tau) + (u/\tau)\chi'(u/\tau)],$$

$$\partial g(u,\tau)/\partial \tau = (-u)^{\alpha + 1}\tau^{-2} (-\chi'(u/\tau)).$$
If $u = u(t)$, then 

$$(d/dt) g(u(t),\tau) = (\partial g(u,\tau)/ \partial u )du/dt + \partial g(u,\tau)/\partial \tau $$  \\
which is positive if $du/dt \ge 0$.  If $u(t) = \tau v - \langle x(t) \rangle $ with $x(t)$ the orbit of a classical particle with Hamiltonian $p^2 + V_1(x)$ then $du/dt = v - \frac{x(t)}{\langle x(t) \rangle }\cdot 2p$ which is positive if $v \ge |2p|$.  This would lead to the useful estimate 

$$( \langle x(t) \rangle - (t+1)v)^{\alpha} \chi(v- \langle x(t) \rangle /(t+1)) \le ( \langle x(0) \rangle - v)^{\alpha} \chi(v- \langle x(0) \rangle )$$
Quantum mechanically we are interested in 

$$ (d/dt) e^{itH} g(A(\tau), \tau) e^{-itH} = e^{itH}\Big( i[H, g(\tau v - \langle x \rangle, \tau) ] + v\partial g(A(\tau),\tau)/\partial u  + \partial g(A(\tau),\tau) /\partial \tau \Big)e^{-itH}.$$
The idea is to show that $$d/dt(\psi_{K,t}, g(A(\tau), \tau) \psi_{K,t}) \ge I(t) $$ where $I(t)$ is integrable on $[0,\infty)$.  Then it would follow that 

$$ (\psi_{K,t}, (-A(\tau))^{\alpha} \chi(A(\tau)/\tau) \psi_{K,t}) \le  (\psi_K, (-A(1))^{\alpha} \chi(A(1)) \psi_K) + \int_0^{\infty}|I(t)|dt$$
We have
$$ i[H, g(\tau v - \langle x \rangle, \tau) ] = - p\cdot (\partial g(A(\tau),\tau)/ \partial u) x/\langle x \rangle - (\partial g(A(\tau),\tau)/ \partial u)( x/\langle x\rangle) \cdot p$$
Let $f(u, \tau) = (-u)^{(\alpha - 1)/2} \tilde \chi(u/\tau)$.

$$ i[H, g(\tau v - \langle x \rangle, \tau) ] +  v\partial g(A(\tau),\tau)/\partial u = - (p \cdot (x/\langle x\rangle)\partial g/\partial u+ \partial g/\partial u \cdot (x/\langle x\rangle) \cdot p) +  v\partial g/\partial u  $$

$$ = -(p \cdot  \hat x f(A(\tau),\tau)^2+  f(A(\tau),\tau)^2\hat x\cdot p) +  v f(A(\tau),\tau)^2$$

$$ = f(v - (\hat x\cdot p + p \cdot\hat x))f$$

where $\hat x = x/\langle x \rangle $.  We have $-(\hat x\cdot p + p \cdot\hat x)) \le p^2/a + a$ so taking $a = v/2$ we have 

$$ i[H, g(\tau v - \langle x \rangle, \tau) ] +  v\partial g(A(\tau),\tau)/\partial u \ge f(v/2 -2 p^2/v) f$$

$$ = 2fV_1f/v + f(v/2 - 2v^{-1} H)f$$

Because $h_1^2(H)H \le (1+\epsilon)^2Kh_1(H)^2$,\ $h_1(H)^2(v/2 - 2v^{-1}H) \ge 0$ and thus

$$ f(v/2 - 2v^{-1} H)f  \ge f h_2 (H)(v/2 - 2v^{-1} H)h_2(H)f \ge -2v^{-1} fh_2(H) H h_2(H)f$$  and since $k_1h_2 = 0$,

$$ d/dt (\psi_{K,t}, g(A(\tau),\tau)\psi_{K,t}) \ge 2v^{-1}(\psi_{K,t}, f^2 V_1(x)\psi_{K,t}) -2v^{-1} (\psi_{K,t} [f,h_2(H)]H[h_2(H), f]\psi_{K,t}). $$
Note that the derivatives of $h_2$ have compact support.  For $n>1$ we have (see \cite{S} but note typos and different definition of $\ad$ - we define $\ad_H(B) = [H,B]$ and $\ad_H^j = (\ad_H)^j$):

$$[h_2(H),f(A(\tau),\tau)] = \sum_{j=1}^{n-1} (-1)^{j+1}h_2^{(j)}(H) \ad_{H}^j(f)/j! + \int_{-\infty}^{\infty} \widehat{ h_2^{(n)}}(w)e^{iwH} R^r_{n,H,f}(w)dw$$ 

$$ = \sum _{j=1}^{n-1}\ad_{H}^j(f) h_2^{(j)}(H)/j!  + \int_{-\infty}^{\infty} \widehat{ h_2^{(n)}}(w) R^l_{n,H,f}(w)e^{iwH}dw.$$
where 

$$R^r_{n,H,f}(w) = \int_0^w (s-w)^{n-1}w^{-n}e^{-isH}\ad_{H}^n(f) e^{isH}ds /(n-1)!/\sqrt{2\pi}. $$ and  

$$ R^l_{n,H,f}(w) = \int_0^w (w-s)^{n-1}w^{-n}e^{isH}\ad_{H}^n(f) e^{-isH}ds /(n-1)!/\sqrt{2\pi}.$$
Note that $h_2^{(j)}(H) \psi_{K,t} = 0$ so that only the remainder terms in the commutators contribute.  We have 

$$\ad^n_{H}(f) = [H,[H,...,[H,f]..] = \sum_{|\gamma| \le n, |\beta|= 2n - |\gamma|} c_{\gamma,\beta}p^{\gamma} D_x^{\beta}f =  \sum_{|\gamma| \le n, |\beta|= 2n - |\gamma|} c'_{\gamma,\beta}D_x^{\beta}f p^{\gamma}$$

$$D_x^{\beta} f(u,\tau) = \sum_{|\gamma| \le |\beta| -1} c_{\gamma}(x) \langle x \rangle^{-|\gamma|} D_u^{|\beta| - |\gamma|}f$$  where $c_{\gamma}(x)$ is a bounded smooth function with bounded derivatives.

 $$(\partial/\partial u)^{\lambda} f(u,\tau) = \sum_{m \le \lambda} c_m (-u)^{(\alpha - 1)/2 -(\lambda -m)} \tau^{-m} \tilde \chi^{(m)}(u/\tau)$$
 In the support of $\tilde \chi^{(m)}(u/\tau)$ for $m > 0$ we have $(-u/\langle x \rangle)^{|\gamma|} \le (2\epsilon v \tau/(\tau(v+\epsilon v))^{|\gamma|} = (2\epsilon/(1+\epsilon))^{|\gamma|}$  while in the support of  $\tilde \chi (u/\tau)$ we have  $0 \le -u/\langle x \rangle \le 1$. 
 It follows that with an $\epsilon$ dependent $c_{\beta}$

 $$ |D_x^{\beta}f| \le c_{\beta}[(-u)^{(\alpha -1)/2 - |\beta|}\tilde \chi(u/\tau) +( \epsilon v) ^{(\alpha - 1)/2 }\tau^{(\alpha - 1)/2 - |\beta| }]$$  We assume $n > (\alpha - 1)/2 + 2$.  Thus since $|\beta| \ge n$ we have $$  |D_x^{\beta}f| \le c_{\beta}'( \epsilon v)^{(\alpha - 1)/2 )}\tau ^{(\alpha - 1)/2 - |\beta|}.$$  We have  
 $$||k_1(H)p^{\gamma} D_x^{\beta}f || \le CK^{|\gamma|/2}(\epsilon v\tau) ^{(\alpha - 1)/2 - |\beta|}$$ and 
 
 $$||k_1(H)\ad^n_{H}(f)|| \le C\sum _{|\beta| \ge n}K^{n + (\alpha -1)/4- |\beta|/2}\tau ^{(\alpha - 1)/2 - |\beta|}$$
 
 We find $\int_{-\infty}^{\infty} |\widehat{h_2^{(n)}}(w)|dw = CK^{-n}$. Then
 
 $$||k_1(H)[f(A(\tau),\tau),h_2(H)]||  \le C\sum _{|\beta| \ge n}K^{(\alpha -1)/4- |\beta|/2}\tau ^{(\alpha - 1)/2 - |\beta|}$$

 We use this for the left commutator in $ (\psi_{K,t}, [f,h_2(H)]H[h_2(H), f]\psi_{K,t})$.  For the term $H[h_2(H), f]\psi_{K,t}$ we use $[h_2(H), f]k_1(H) = - \int_{-\infty}^{\infty} \widehat{ h_2^{(n)}}(w) R^l_{n,H,f}(w)e^{iwH}dwk_1(H)$
so that
$$||H[h_2(H), f]k_1(H)||  \le C\sum _{|\beta| \ge n}K^{1 +(\alpha -1)/4 - |\beta|/2}\tau^{(\alpha - 1)/2 - |\beta|} + C\sum _{|\beta| \ge n+1}K^{1 + (\alpha - 1)/4 - |\beta|/2}\tau ^{(\alpha - 1)/2 - |\beta|} .$$  It follows that with $K \ge 1$
$$v^{-1}||k_1(H)[f,h_2(H)]H[h_2(H),f]k_1(H)|| \le C K^{(\alpha +2 - 2n)/2} \tau^{\alpha - 1 - 2n}.$$ Since we have already taken $2n > \alpha +3$ we have

 $$ (\psi_{K,t}, (-A(\tau))^{\alpha} \chi(A(\tau)/\tau) \psi_{K,t}) \le  (\psi_K, (-A(1))^{\alpha} \chi(A(1)) \psi_K) + v^{-1}\int_0^{\infty}(\psi_{K,t}, |V_1(x)|f^2\psi_{K,t} )dt + C'(K)$$
 where $\lim_ {K\to \infty} C'(K) = 0$.  Since $V_1$ is bounded with compact support $|V_1(x)|f^2 \le c_N (v\tau)^{-N}$ for any $N$ and thus
  $$ (\psi_{K,t}, (-A(\tau))^{\alpha} \chi(A(\tau)/\tau) \psi_{K,t}) \le  (\psi_K, (-A(1))^{\alpha} \chi(A(1)) \psi_K) + C''(K) \le ||\langle x \rangle ^{\alpha/2}\psi_K||^2 + C''(K)$$ 
  where $\lim_ {K\to \infty} C''(K) = 0$.
  Since in the support of $ \chi(A(\tau)/\tau), \langle x \rangle -v\tau \ge \epsilon \langle x \rangle/(1+\epsilon)$ we have
  $$ (\psi_{K,t}, \langle x \rangle^{\alpha} \chi(A(\tau)/\tau) \psi_{K,t}) \le (1+ \epsilon^{-1})^{\alpha}||\langle x \rangle ^{\alpha/2}\psi_K||^2 + C'''(K).$$ In addition $\chi(A(\tau)/\tau)) = 1$ if $\langle x \rangle \ge (1+2\epsilon)v\tau$ which gives Proposition \ref{propest}.

  \end{proof}
  
  \begin{proof} of Theorem \ref{feelingV_L-2}
  
  Now back to what we want to bound:  $$\int_0^t ||V_{2,L}(x)e^{-isH}\psi ||ds \le ||V_{2,L}||_{\infty}\Big ( t || k_2(H)\psi|| +L^{-\alpha/2}\int_0^t ||1_{\{\langle x \rangle \ge L\}}\langle x \rangle ^{\alpha/2}e^{-isH} \psi_K|| ds \Big).$$
  We have used $1-k_1 \le k_2$.    Since $\psi \in \mathcal{S}(\mathbb R)$,  $||k_2(H)\psi|| \le c_N K^{-N}$.  
  
  We have
  
  $$||1_{\{\langle x \rangle \ge L\}}\langle x \rangle ^{\alpha/2}e^{-isH}\psi_K|| 
  \le ||1_{\{\langle x \rangle \ge (1+ 2\epsilon) v (s+1)\}}\langle x \rangle ^{\alpha/2}e^{-isH} \psi_K|| $$
  where we take $(1+ 2\epsilon) v(t+1)\le L$ and $0 \le s \le t$.   It follows from Proposition \ref{propest} that 
   $$||1_{\{\langle x \rangle \ge L\}}\langle x \rangle ^{\alpha/2}e^{-isH}k_1(H) \psi|| \le (1 + \epsilon^{-1})^{\alpha} ||\langle x \rangle ^{\alpha/2}e^{-isH}k_1(H) \psi|| + C_{\alpha}(K)$$  where $\lim _{K \to \infty} C_{\alpha}(K) =0$.
  
  Thus we have 
  $$\int_0^t ||V_{2,L}(x)e^{-isH}\psi ||ds \le t(c_NK^{-N} + cL^{-\alpha/2}(||\langle x \rangle ^{\alpha/2}k_1(H)\psi|| +1).$$
  We take $K= L^{2\delta}$ for $0 < \delta << 1$ and thus $t \le (1+\epsilon)^{-2} L/ \sqrt {4K} = cL^{1-\delta}$.
 Since $\alpha$ is arbitrary we have 
$$\int_0^t ||V_{2,L}(x)e^{-isH}\psi ||ds \le c_nL^{-n}$$
as long as $t\le cL^{1-\delta}$ and we can prove that $||\langle x \rangle ^{\alpha/2}k_1(H)\psi||$ is bounded as $K\to \infty$.  To deal with  $||\langle x \rangle ^{\alpha/2}k_1(H)\psi||$ write the square $$(\psi,k_1(H)\langle x \rangle ^{\alpha}k_1(H)\psi) = (\psi, k_1(H)^2\langle x \rangle ^{\alpha}\psi) + (\psi,k_1(H)[\langle x \rangle ^{\alpha},k_1(H)]\psi).$$  We use 

$$[k_1(H),f] = \sum_{j=1}^{n-1}(-1)^{j+1} k_1^{(j)}(H) \ad_{H}^j(f)/j! + \int_{-\infty}^{\infty} \widehat{ k_1^{(n)}}(w)e^{iwH} R^r_{n,H,f}(w)dw$$ where $f = \langle x \rangle ^{\alpha}$.

We have 
$$\ad^j_{H}(\langle x \rangle^{\alpha} ) = [H,[H,...,[H,\langle x \rangle ^{\alpha}]..] = \sum_{|\gamma| \le j, |\beta|= 2j - |\gamma|} c_{\gamma,\beta}p^{\gamma} D_x^{\beta}\langle x \rangle ^{\alpha}$$ and thus

$$|| k_1^{(j)}(H)\ad^j_{H}(\langle x \rangle^{\alpha} )\psi|| \le CK^{-j}K^{j/2}||\langle x \rangle^{\alpha} \psi||.$$  Finally  $$\int _{-\infty}^{\infty}|| k_1(H)R^r_{n,H,f}|| \ |\widehat{k_1^{(n)}}(w)| dw \le CK^{-n}K^{n/2}$$
if $n\ge \alpha$.
This gives Theorem \ref{feelingV_L-2}.
\end{proof}

\section{Appendix: Lavine's bound}
\newcommand{\jap}[1]{{\langle#1\rangle}}

In this appendix we present Lavine's bound on the time evolution of resonant states. 

\begin{theorem} Let $H=-d^2/dx^2 + V$ acting in $L^2(\bbR)$, where $V$ satisfies the hypothesis of Section 2 with $\supp(V)\subseteq [-r,r]$. Suppose that $k_0$ is a resonance and let $\psi$ be the corresponding outgoing solution of 
\begin{equation}\label{req}
-\psi''(x) + (V(x) - k_0^2)\psi(x) = 0.
\end{equation}
Recall that outgoing means that $\psi(x)$ is a multiple of $e^{-ikx}$ for $x\le -r$ and a multiple of $e^{ikx}$ for $x\ge r$. Write $k_0 = \sigma_0 - i\epsilon_0$ and $k_0^2 = \lambda_0-i\delta_0$ with $\sigma_0,\epsilon_0,\delta_0 > 0$. Let $\chi=\chi_{[-r,r]}$ be the indicator function for $[-r,r]$ and define $\varphi=\chi\psi$. Then
\begin{equation}\label{lavineest}
\left|\langle{\varphi,\left(e^{-itH} - e^{-it\lambda_0-\delta_0|t|}\right)\varphi}\rangle\right| \le C(k_0)\|\varphi\|^2
\end{equation}
where
$$
C(k_0)=\left (\frac{1}{5}\log(1+ (\frac{\sigma_0}{2\epsilon_0})^2) + 1\right )\frac{6\epsilon_0}{\sigma_0}
$$
\end{theorem}

\begin{remark} The constant $C$ is $O(({\epsilon_0}/{\sigma_0})\log(\sigma_0/\epsilon_0))$ for small ${\epsilon_0}/{\sigma_0}$, and is independent of $V$, $r$ and $t$. No spectral cutoff is required for $\varphi$, so the bound implies that $\varphi$ is almost orthogonal to bound states if $\epsilon_0$ is small. Lavine's result is for the half line. To generalize this to the whole line we rewrite his bounds in terms of interior and exterior Dirichlet to Neumann maps and use that these maps commute.
\end{remark}

Following Lavine, we begin the proof of this theorem by introducing the Lorenzian 
distribution given by
$$
d\mu_L(\lambda) ={{1}\over{\pi}}\Im (\lambda_0-i\delta_0-\lambda)^{-1} d\lambda.
$$
The Fourier transform of this distribution is $e^{-it\lambda_0-\delta_0|t|}$.
Thus, if $\varphi\in L^2(\bbR)$ has spectral measure $d\mu_\varphi(\lambda)$,
we find that
\begin{equation}\label{ftformula}
\langle{\varphi,\left(e^{-itH} - e^{-it\lambda_0-\delta_0|t|}\right)\varphi}\rangle
= \int_\bbR e^{-it\lambda} \left(d\mu_\varphi(\lambda)-\|\varphi\|^2d\mu_L(\lambda)\right).
\end{equation}
This means that we will be able to approximate the time evolution of $\varphi$ 
if the spectral measure $\mu_\varphi$ is well approximated by $\|\varphi\|^2\mu_L$. In fact, it is enough to make this approximation on an interval $I$ outside which $\mu_L(\lambda)$ is small. 
\begin{proposition}\label{lavineprop}
Suppose that for some interval $I\subset \bbR$ we have
\begin{equation}\label{diffest}
\int_I \left|d\mu_\varphi(\lambda) - \|\varphi\|^2d\mu_L(\lambda)\right| \le \epsilon_1\|\varphi\|^2
\end{equation}
and
\begin{equation}\label{lorentzest}
\int_{\bbR\backslash I} d\mu_L(\lambda) \le \epsilon_2.
\end{equation}
Then
$$
\left|\langle{\varphi,\left(e^{-itH} - e^{-it\lambda_0-\delta_0|t|}\right)\varphi}\rangle\right|
\le 2(\epsilon_1+\epsilon_2)\|\varphi\|^2
$$
\end{proposition}
\begin{proof}
From \eqref{ftformula} we find
\begin{align*}
\left|\jap{\varphi,\left(e^{-itH} - e^{-it\lambda_0-\delta_0|t|}\right)\varphi}\right|
&\le \int_I \left|\left(d\mu_\varphi(\lambda)-\|\varphi\|^2d\mu_L(\lambda)\right)\right|
+ \int_{\bbR\backslash I}\|\varphi\|^2d\mu_L(\lambda)  + \int_{\bbR\backslash I}d\mu_\varphi(\lambda) \\
&\le (\epsilon_1 + \epsilon_2)\|\varphi\|^2 + \int_{\bbR\backslash I}d\mu_\varphi(\lambda)\\
\end{align*}
Since the measures $d\mu_\varphi$ and  $\|\varphi\|^2d\mu_L$ both have total mass $\|\varphi\|^2$,
\begin{align*}
\int_{\bbR\backslash I}d\mu_\varphi(\lambda)
&=\|\varphi\|^2-\int_{I}d\mu_\varphi(\lambda)\\
&=\|\varphi\|^2-\|\varphi\|^2\int_{I}d\mu_L(\lambda) + \int_{I}\left(\|\varphi\|^2d\mu_L(\lambda)-d\mu_\varphi(\lambda)\right)\\
&\le \|\varphi\|^2\int_{\bbR\backslash I}d\mu_L(\lambda) + \int_{I}\left|d\mu_\varphi(\lambda)-\|\varphi\|^2d\mu_L(\lambda)\right|\cr
&\le (\epsilon_1+\epsilon_2)\|\varphi\|^2.\\
\end{align*}
This completes the proof.
\end{proof}

Recall that the resolvent $R(k)=(H-k^2)^{-1}$, initially defined for $\Im(k)>0$, has an integral kernel given by the\ \GF\ $G(x,y;k)$. This representation defines the limit $\jap{\varphi, R(k+i0)\varphi}$ onto the real line (and the continuation beyond). We denote the continuation by $\jap{\varphi, R(k)\varphi}$ even when the operator $R(k)$ is not defined. With our assumptions on $V$, the spectral measure $d\mu_\varphi(\lambda)$ is absolutely continuous for $\lambda\in[0,\infty)$ with density $\pi^{-1}\Im(\jap{\varphi, (H-\lambda - i0)^{-1}\varphi})$. Thus the crucial quantity to be estimated in \eqref{diffest} can be written
\begin{align*}
\int_I \left|d\mu_\varphi(\lambda) - \|\varphi\|^2d\mu_L(\lambda)\right|
&=
{{1}\over{\pi}}\int_I\left|\Im\left[
\jap{\varphi, (H-\lambda - i0)^{-1} \varphi} - \|\varphi\|^2 (\lambda_0-i\delta_0-\lambda)^{-1}
\right]\right|d\lambda\\
&={{1}\over{\pi}}\int_{\sqrt{I}}\left|\Im
\jap{\varphi,\left[R(k) - {{1}\over{k_0^2-k^2}}\right]\varphi}
\right|2kdk\numberthis \label{resolventest}\\
\end{align*}
Here we are assuming that $I\subseteq [0,\infty)$.
We now compute another expression for the integrand. 
\begin{proposition} Let $\psi$ be the outgoing solution to \eqref{req}. For $\Im(k)>0$ and $\chi\in C_0^\infty(\bbR)$ we have
\begin{equation}\label{coe2}
R(k)\chi\psi = {{1}\over{k_0^2-k^2}}\chi\psi - {{1}\over{k_0^2-k^2}}R(k)[H,\chi]\psi,
\end{equation}
and if $\chi_1\in C_0^\infty$ with $\chi_1[H,\chi]=0$ and $k$ is real, then
\begin{equation}\label{coe3}
\jap{\chi_1\psi,\left[R(k) - {{1}\over{k_0^2-k^2}}\right]\chi\psi}
={{1}\over{\left|k_0^2-k^2\right|^2}}\jap{[H,\chi_1]\psi,R(k)[H,\chi]\psi}
\end{equation}
\end{proposition}
\begin{proof}
Equation \eqref{coe2} follows from
$$
(H-k^2)\chi\psi = (k_0^2-k^2)\chi\psi +[H,\chi]\psi.
$$
To prove \eqref{coe3} we use \eqref{coe2} (twice) and $\chi_1[H,\chi]=0$ to write, for $\Im(k)>0$,
\begin{align*}
\jap{\chi_1\psi,\left[R(k) - {{1}\over{k_0^2-k^2}}\right]\chi\psi}
&={{-1}\over{k_0^2 - k^2}}\jap{\chi_1\psi,R(k)[H,\chi]\psi}\\
&={{-1}\over{k_0^2 - k^2}}\jap{R(-\overline k)\chi_1\psi,[H,\chi]\psi}\\
&={{1}\over{(k_0^2 - k^2)(\overline k_0^2-k^2)}}
\jap{-\chi_1\psi+R(-\overline k)[H,\chi_1]\psi,[H,\chi]\psi}\\
&={{1}\over{(k_0^2 - k^2)(\overline k_0^2-k^2)}}
\jap{[H,\chi_1]\psi,R(k)[H,\chi]\psi}\\
\end{align*}
Here we used that $R(k)^* = R(-\overline k)$. Now we take the limit on the real axis.
\end{proof}
We will use the following integration by parts formula
\begin{proposition} Let $\psi$ be the outgoing solution to \eqref{req} and $\varphi=\chi_{[-r,r]}\psi$. Then
\begin{equation}
\left\|\begin{bmatrix}\psi(-r)\\\psi(r)\\\end{bmatrix}\right\|^2 = 2\epsilon_0\|\varphi\|^2
\end{equation}
\end{proposition}
\begin{proof}
\begin{align*}
2i\delta_0\|\varphi\|^2
&=2i\delta_0\|\psi\|^2_{L^2([-r,r])}\\
&= \jap{(\lambda_0-i\delta_0)\psi,\psi}_{L^2([-r,r])} - \jap{\psi,(\lambda_0-i\delta_0)\psi}_{L^2([-r,r])}\\
&= \jap{(-d^2/dx^2+V)\psi,\psi}_{L^2([-r,r])} - \jap{\psi,(-d^2/dx^2+V)\psi}_{L^2([-r,r])}\\
&= -\overline\psi'(x)\psi(x)\Big|_{x=-r}^r + \overline\psi(x)\psi'(x)\Big|_{x=-r}^r\\
\end{align*}
This implies
\begin{equation}
\delta_0\|\varphi\|^2 = \Im\left(\left\langle\begin{bmatrix}\psi(-r)\\\psi(r)\\\end{bmatrix},\begin{bmatrix}-\psi'(-r)\\\phantom{-}\psi'(r)\\\end{bmatrix}\right\rangle\right).
\end{equation}
Since $\psi$ is outgoing, $\begin{bmatrix}-\psi'(-r)\\\phantom{-}\psi'(r)\\\end{bmatrix}=ik_0\begin{bmatrix}\psi(-r)\\\psi(r)\\\end{bmatrix}$. Since $k_0=\sigma_0-i\epsilon_0$  and $\delta_0=2\epsilon_0\sigma_0$ this gives the result.
\end{proof}

We now introduce the interior and exterior Dirichlet to Neumann maps. Given $\begin{bmatrix}a\\b\\\end{bmatrix}\in\bbC^2$, solve $-\psi'' + (V-k^2)\psi = 0$ on $[-r,r]$ with $\psi(-r)=a$ and $\psi(r)=b$. This is possible as long as $k^2$ is not a Dirichlet eigenvalue of $-d^2/dx^2 + V$. Then the interior Dirichlet to Neumann map is defined to be
$$
\Lambda_k\begin{bmatrix}a\\b\\\end{bmatrix}=\begin{bmatrix}-\psi'(-r)\\\psi'(r)\\\end{bmatrix}.
$$
When $k\in\bbR$, $\Lambda_k$ is a real symmetric $2\times 2$ matrix (when it is defined). 

The exterior Dirichlet to Neumann map is defined similarly, except we now find the outgoing solution to $-\psi''-k^2\psi=0$ on $(-\infty,-r] \cup [r,\infty)$, again with $\psi(-r)=a$ and $\psi(r)=b$. The solution is simply $ae^{-ik(x+r)}$ on $(-\infty,-r]$ and $be^{ik(x-r)}$ on $[r,\infty)$ so the exterior Dirichlet to Neumann map is
$$
\Omega_k\begin{bmatrix}a\\b\\\end{bmatrix}=\begin{bmatrix}-\psi'(-r)\\\psi'(r)\\\end{bmatrix} = ik\begin{bmatrix}a\\b\\\end{bmatrix}.
$$
So we see that $\Omega_k = ikI$ where $I$ is the $2\times 2$ identity matrix. 

Now let $r' > r$ and introduce the matrices
$$
{\bf G}(r,r',k) = \begin{bmatrix}G(-r,-r';k)&G(-r,r';k)\\G(r,-r';k)&G(r,r';k)\\\end{bmatrix}
$$
which are the restrictions of the resolvent to ``spheres''. Since the point $-r'$ is outside the interval $[-r,r]$, the functions $G(x,-r';k)$ and $G(x,r';k)$ solve the equation $(-d^2/dx^2+V(x)-k^2)G(x,\pm r';k)=0$ in the interval $[-r,r]$. This implies that
$$
\begin{bmatrix}-G_x(-r,-r';k)&-G_x(-r,r';k)\\G_x(r,-r';k)&G_x(r,r';k)\\\end{bmatrix} = \Lambda_k {\bf G}(r,r',k).
$$
Similarly
$$
\begin{bmatrix}-G_y(-r,-r';k)& G_y(-r,r';k)\\-G_y(r,-r';k)&G_y(r,r';k)\\\end{bmatrix} =  {\bf G}(r,r',k)\Omega_k^T ={\bf G}(r,r',k)\Omega_k
$$
and
$$
\begin{bmatrix}G_{x,y}(-r,-r';k)&-G_{x,y}(-r,r';k)\\-G_{x,y}(r,-r';k)&G_{x,y}(r,r';k)\\\end{bmatrix} = \Lambda_k {\bf G}(r,r',k)\Omega_k.
$$

Now we take the limit as $r'\rightarrow r$ and define 
$${\bf G}(k)= \lim_{r'\rightarrow r}{\bf G}(r,r',k).$$

\begin{proposition} The matrix ${\bf G}(k)$ has the representation
$$
{\bf G}(k) = (\Lambda_k - \Omega_k)^{-1}
$$
\end{proposition}
\begin{proof}
$\Lambda_k {\bf G}(r,r';k)$ is given by the formula above. To compute $\Omega_k{\bf G}(r,r';k)$ introduce the exterior Dirichlet\ \GF\ $G_0(x,y,k)$ on $(-\infty,-r] \cup [r,\infty)$. This can be computed explicitly. 
Then $G(x,\pm r';k)-G_0(x,\pm r';k)$ is an outgoing solution to $(-d^2/dx^2-k^2)G(x,\pm r';k)=0$ for $x\in(-\infty,-r] \cup [r,\infty)$. This implies that
$$
\Omega_k {\bf G}(r,r';k) = \begin{bmatrix}-G_x(-r,-r';k)+G_{0,x}(-r,-r';k)&-G_x(-r,r';k)+G_{0,x}(-r,r';k)\\G_x(r,-r';k)-G_{0,x}(r,-r';k)&G_x(r,r';k)-G_{0,x}(r,r';k)\\\end{bmatrix}
$$
so that
\begin{equation}\label{dirgreen}
(\Lambda_k - \Omega_k){\bf G}(r,r';k) = \begin{bmatrix}-G_{0,x}(-r,-r';k)&-G_{0,x}(-r,r';k)\\G_{0,x}(r,-r';k)& G_{0,x}(r,r';k)\\\end{bmatrix}
\end{equation}
Since $(-\infty,-r] \cup [r,\infty)$ is disconnected, $G_{0,x}(r,-r';k)=G_{0,x}(-r,r';k)=0$ To compute $G_{0,x}(r,r';k)$ let $f(x)$ be an outgoing solution of $(-d^2/dx^2 - k^2)f=0$ on $[r,\infty)$. Explicitly $f(x)=f(r)e^{ik(x-r)}$. Then, since $(-d^2/dx^2 - k^2)G_0(x,r';k)=\delta(r-r')$, we have that for $R>r'$
\begin{align*}
f(r') &= \int_r^R \big((-d^2/dx^2 - k^2)G_0(x,r';k)\big) f(x)dx\\
&=-G_{0,x}(x,r';k)f(x)\Big|_{x=r}^R + G_{0}(x,r';k)f_x(x)\Big|_{x=r}^R + \int_r^R G_0(x,r';k)\big((-d^2/dx^2 - k^2)f(x)\big) dx\\
&=-G_{0,x}(R,r';k)f(R) + G_{0,x}(r,r';k)f(r)+G_{0}(R,r';k)f_x(R)-G_{0}(r,r';k)f_x(r)\\
&=-ikG_{0}(R,r';k)f(R) + G_{0,x}(r,r';k)f(r)+G_{0}(R,r';k)ikf(R) - 0\\
&=G_{0,x}(r,r';k)f(r)\\
\end{align*}
Here we used that $G_0(x,r';k)$ and $f(x)$ are both outgoing at $x=R$, and $G_{0}(r,r';k)=0$. Now take $r'\rightarrow r$ to conclude $\lim_{r'\rightarrow r}G_{0,x}(r,r';k)=1$. Similary $\lim_{r'\rightarrow r}G_{0,x}(-r,-r';k)=-1$.
So the right side of \eqref{dirgreen} converges to $I$ as $r'\rightarrow r$. This completes the proof.
\end{proof}
\begin{remark}
This proposition and the preceeding formulas can also be proven using the explicit representation for the\ \GF\ in terms of $\psil$ and $\psir$. The above proof has the advantage that it generalizes to higher dimensions. However, in higher dimensions $\Omega_k$ and $\Lambda_k$ no longer commute, unless, for example, $V$ is radial.
\end{remark}

Now we return to \eqref{coe3} and rewrite the right side. We have 
\[
\jap{[H,\chi_1]\psi,R(k)[H,\chi]\psi} = \jap{(D_x\chi_1' + \chi_1'D_x)\psi,R(k)(D_x\chi' + \chi'D_x)\psi}
\]
Here $D_x$ is the operator of differentiation by $x$. Let $r'>r$ and choose $\chi_1$ to be a smooth cutoff to $[-r,r]$ and $\chi$ a smooth cutoff to $[-r',r']$ chosen so that the derivatives $\chi_1'$ and $\chi'$ have disjoint support. Then $\chi_1'(x) = \delta_{-r}(x) - \delta_r(x)$ and $\chi'(x) = \delta_{-r'}(x) - \delta_{r'}(x)$ where $\delta_{-r}(x), \delta_r(x), \delta_{-r'}(x),\delta_{r'}(x)$ are approximate delta functions with disjoint support. Now we write the right side as a double integral with\ \GF s in the integrand. Then integrate by parts, until the derivatives are hitting either the\ \GF\ or $\psi$. Then we let take the limit as the approximate delta functions become exact delta functions. Finally we let $r'\rightarrow r$. A typical term in this calculation looks like
\begin{align*}
&\jap{D_x(\delta_{-r}(x) - \delta_r(x))\psi,R(k)(\delta_{-r'}(x) - \delta_{r'}(x))D_x\psi }\\
&=-\int\int (\delta_{-r}(x) - \delta_r(x))\overline\psi(x) G_x(x,y;k)(\delta_{-r'}(y) - \delta_{r'}(y))D_y\psi(y) dy dx\\
&=\left\langle\begin{bmatrix}\psi(-r)\\\psi(r)\\\end{bmatrix}, \Lambda_k{\bf G}(k)\begin{bmatrix}\psi'(-r)\\-\psi'(r)\\\end{bmatrix}\right\rangle\\
&=\left\langle\begin{bmatrix}\psi(-r)\\\psi(r)\\\end{bmatrix}, -\Lambda_k{\bf G}(k)\Omega_{k_0}\begin{bmatrix}\psi(-r)\\\psi(r)\\\end{bmatrix}\right\rangle\\
\end{align*}
In this way we arrive at
\begin{align*}
{{1}\over{\left|k_0^2-k^2\right|^2}}&\jap{[H,\chi_1]\psi,R(k)[H,\chi]\psi}\\
&={{1}\over{\left|k_0^2-k^2\right|^2}}\left\langle\begin{bmatrix}\psi(-r)\\\psi(r)\\\end{bmatrix},(\Lambda_k-\Omega_{k_0}^*)(\Lambda_k-\Omega_k)^{-1}(\Omega_k-\Omega_{k_0})\begin{bmatrix}\psi(-r)\\\psi(r)\\\end{bmatrix}\right\rangle
\end{align*}
Recall that $\Omega_k=ik$. This leads to
\begin{align*}
\|(\Lambda_k-\Omega_{k_0}^*)(\Lambda_k-\Omega_k)^{-1}(\Omega_k-\Omega_{k_0})\| &= 
\|(\Omega_k-\Omega_{k_0}) +(\Omega_k-\Omega_{k_0}^*)(\Lambda_k-\Omega_k)^{-1}(\Omega_k-\Omega_{k_0})\|\\
&\le |k-k_0|\left(1 + \frac{|k+\overline k_0|}{|k|}\right)\\
\end{align*}
So returning to \eqref{resolventest} with $I =[(\sigma_0/2)^2, (3\sigma_0/2)^2]$, $\sqrt{I}=[\sigma_0/2, 3\sigma_0/2]$ we find
\begin{align*}
\int_I \left|d\mu_\varphi(\lambda) - \|\varphi\|^2\mu_L(\lambda)\right|
&={{1}\over{\pi}}\int_{I}\left|\Im
\jap{\varphi,\left[R(k) - {{1}\over{k_0^2-k^2}}\right]\varphi}
\right|dk^2\\
&\le {{1}\over{\pi}}\int_I \frac{|k-k_0|}{|k_0^2-k^2|^2}\left(1 + \frac{|k+\overline k_0|}{k}\right)\left\|\begin{bmatrix}\psi(-r)\\\psi(r)\\\end{bmatrix}\right\|^2dk^2\\
&\le  {{4\epsilon_0}\over{\pi}}\int_{\sqrt{I}}\frac{k+|k+k_0|}{|k_0+k|^2}\frac{1}{|k-k_0|}dk\; \|\varphi\|^2\\
\end{align*}
We have
$$\frac{4\epsilon_0}{\pi} \int_{\sqrt{I}} \frac{k + |k+k_0|}{|k_0 + k|^2 |k-k_0|} dk \le \frac{4\epsilon_0}{\pi} \int_{\sqrt{I}} \frac{k + |k+\sigma_0|}{|\sigma_0 + k|^2 |k-k_0|} dk$$

The derivative of $\frac{2k+\sigma_0}{(k+\sigma_0)^2} $ is negative so in this term we replace $k$ by $\sigma_0/2$. Thus 

$$\frac{4\epsilon_0}{\pi} \int_{\sqrt{I}} \frac{k + |k+k_0|}{|k_0 + k|^2 |k-k_0|} dk \le \frac{32}{9\pi}(\epsilon_0/\sigma_0)\int_{\sqrt{I}} \frac{dk}{|k-k_0|}  \le \frac{32}{9\pi}(\epsilon_0/\sigma_0)\int_{-1/2}^{1/2} \frac{dx}{\sqrt{x^2 + (\epsilon_0/\sigma_0)^2}}$$
$$ \le  \frac{64}{9\pi}(\epsilon_0/\sigma_0)\log \left( \frac{\sigma_0}{2\epsilon_0} + \sqrt {1 + (\frac{\sigma_0}{2\epsilon_0})^2}\right)$$

The other estimate is for $\int_{\mathbb{R} \setminus I}d\mu_L(\lambda)$.  We first estimate the contribution from $[0,\infty)\setminus I$ which is 

$$\pi^{-1} \int_{[0,\infty)\setminus {\sqrt{I}}}|\text {Im} \frac{1}{k^2-k_0^2}|2kdk = \pi^{-1} \int_{[0,\infty)\setminus {\sqrt{I}}}|\text {Im} (\frac{1}{k-k_0} + \frac{1}{k+k_0})|dk$$
$$ = \epsilon_0\pi^{-1} \int_{[0,\infty)\setminus {\sqrt{I}}}(\frac{1}{|k-k_0|^2} - \frac{1}{|k+k_0|^2})dk \le \epsilon_0\pi^{-1} \int_{[0,\infty)\setminus {\sqrt{I}}}\frac{1}{|k-k_0|^2} dk$$
$$ \le  \epsilon_0\pi^{-1}   \int _ {[0,\infty)\setminus {\sqrt{I}}}\frac {1}{(k-\sigma_0)^2 }dk = \frac{3 \epsilon_0}{\pi \sigma_0}.$$  

Next we integrate over $(-\infty,0)$.

$$\pi^{-1} \int_{-\infty}^0|\text{Im} (\lambda - k_0^2)^{-1}|d\lambda = \pi^{-1} \int_0^{\infty}|\text{Im} (-k^2 - k_0^2)^{-1}|2kdk = \pi^{-1} \int_0^{\infty}|\text{Im}( \frac{1}{k+ik_0} + \frac{1}{k-ik_0})|dk $$
$$ = \sigma_0\pi^{-1} \int_0^{\infty}( \frac{1}{|k-ik_0|^2} - \frac{1}{|k+ik_0|^2})dk = 4 \sigma_0\epsilon_0\pi^{-1} \int_0^{\infty}( \frac{k}{|k-ik_0|^2|k+ik_0|^2} dk$$
$$\le 4 \sigma_0\epsilon_0\pi^{-1} \int_0^{1+\epsilon_0}\frac{1}{2\sigma_0} \frac{1}{(k-\epsilon_0)^2 + \sigma_0^2} dk + 4 \sigma_0\epsilon_0\pi^{-1} \int_{1+\epsilon_0}^{\infty}\frac{1}{\sigma_0} \frac{1}{(k-\epsilon_0)^2 + \sigma_0^2} dk $$
$$\le \frac{4\epsilon_0}{\sigma_0\pi}( \pi/2 + \pi/2) = \frac{4\epsilon_0}{\sigma_0}$$

Summing we get for the total error bound

$$\left (\frac{64}{9\pi}\log \left( \frac{\sigma_0}{2\epsilon_0} + \sqrt {1 + (\frac{\sigma_0}{2\epsilon_0})^2}\right) + (4+3/\pi) \right)\frac{\epsilon_0}{\sigma_0}$$
$$ <\left (\frac{1}{5}\log(1+ (\frac{\sigma_0}{2\epsilon_0})^2) + 1\right )\frac{6\epsilon_0}{\sigma_0}$$
This completes the proof of Lavine's estimate.

\def\twomat#1#2{\begin{bmatrix} #1 \\ #2 \\\end{bmatrix}}
\section{Appendix: The function $f_2$ for small coupling}
When $H$ is a Hamiltonian of the form $-d^2/dx^2 + V_1(x) +  \mu V_{2}(x-L)$ the function $f_2(k,\mu)$ defined in (\ref{eqn:f}) will depend on $\mu$. In this appendix we establish the small $\mu$ behaviour of this function. Since $V_1$ is not involved in the definition of $f_2$ we drop the subscript on $V_2$.
\begin{proposition}\label{f2smallmu}
Let $V(x) = V_{0}(x) + \sum_{i=1}^N\alpha_i\delta(x-x_i)$ where $0=x_1<x_2<\cdots<x_N=x_f$, $V_{0}(x)$ is continuous and  $\supp(V_0)\subseteq [0,x_f]$.  Suppose that $\widehat V(-2ik) \ne 0$ {where $\widehat V (k) = \int V(x) e^{-ikx} dx$}.
Then the function $f_2(k,\mu)$ defined in (\ref{eqn:f}) sastisfies
\begin{align*}
\mu f_2(k,\mu ) &=\frac{ -2ik}{\widehat V(-2ik) } +
\mu\left(\frac{\widehat V(0)}{\widehat V(-2ik)}\right.\\
&+\left.\frac{1}{2ik\widehat V(-2ik)^2}\int_0^{x_f}\int_{t_1}^{x_f}V(t_1)V(t_2)(-e^{ikt_1})\sin((t_1-t_2)k)e^{ikt_2}dt_1dt_2\right) + O(\mu^2)
\end{align*}
where the $O(\mu^2)$ term is uniform for $k$ in a bounded set bounded away from zero.
\end{proposition}

To establish this we let $\psi_2(x,k,\mu)$ be the solution of $-\psi'' +\mu V(x)\psi = k^2\psi$  with $\psi(x)=e^{ikx}$ for $x\ge x_N$. We must compute the expression in  (\ref{eqn:f}). 
The vector $\Psi = \twomat{\psi(x)}{\psi'(x)}$ satisfies the first order system
$$
\Psi' =\twomat{0&1}{V_2(x) - k^2&0} \Psi
$$
away from the $x_i$. At the points $x_i$ we have
\be\label{deltajump}
\Psi(x_i+) = \twomat{1&0}{\alpha_i&1}\Psi(x_i-).
\ee
We now write 
$$
\psi_2(x,k,\mu) = a(x,k,\mu)e^{ikx} + b(x,k,\mu)e^{-ikx}
$$
where
$$
a'(x,k,\mu)e^{ikx} + b'(x,k,\mu)e^{-ikx} = 0.
$$
and define
$$X(x)=X(x,k,\mu)=\twomat{ a(x,k,\mu)}{b(x,k,\mu)}.$$ 
Then
$$
 \Psi(x) = \twomat{e^{ikx} &e^{-ikx} }{ike^{ikx} &-ike^{-ikx}}X(x).
$$
and a short calculation shows that  away from the points $x_i$, $X$
satisfies the first order system
$$
X'(x) = \frac{V(x)}{2ik}A(x,k)X(x).
$$
where
$$
A(x,k) =\twomat{1&e^{-2ikx}}{-e^{2ikx}&-1}
$$
The change in $X$ as $x$ passes a point $x_i$ is given by
\be\label{onestep}
X(x_i-) = \left(I - \frac{\alpha_i}{2ik}A(x_i,k)\right)X(x_i+).
\ee
The solution $X$  that satisfies the final condition $X_f=X(x_f)=\twomat{1}{0}$ solves  the integral equation
\begin{align}\label{inteq}
X(x) = X_f - \int_{x}^{x_{f}}\frac{\mu V(t)}{2ik}A(t,k)X(t)dt.
\end{align}
provided that for a piecewise continuous vector function $Y(x)$ that may have jump discontinuities at the points $x_i$, we  interpret
$$
\int_{x_i-}^{x_i+} \delta(x-x_i) Y(x)dx = Y(x_i+).
$$
The solution $X(x)=\twomat{ a(x,k,\mu)}{b(x,k,\mu)}$ to \eqref{inteq} then determines $f_2$ via
$$
f_2(k,\mu) = -\frac{a(0-,k,\mu)}{b(0-,k,\mu)}.
$$

We will need an a priori bound on $X(x)$. Starting with $\|X(x_N+)\|=\|X_f\| = 1$ we use \eqref{onestep} to estimate
\begin{align*}
\|X(x_N-)\| &\le \left(1+\frac{\mu |\alpha_N| M(k)}{2|k|}\right)\|X(x_N+)\| \\
&\le \exp\left({\frac{\mu |\alpha_N| M(k)}{2|k|}}\right)
\end{align*}
where $M(k) = \sup_{[x\in[0,x_f]}\|A(k,x)\| = 2\cosh(x_f\Im(k))$.
Then, using Gronwall's inequality, it is not hard to see that
$$
\|X(x_{N-1}+)\|\le \exp\left({\frac{\mu \|V_0\|_\infty(x_{N}-x_{N-1}) M(k)}{2|k|}}\right)\|X(x_N-)\|.
$$
Continuing like this, we arrive at the bound
\be\label{apriori}
\sup_{x\in[0,x_f]}\|X(x)\| \le \exp\left({\frac{\mu M(k) (\|V_0\|_\infty x_f+ \sum|\alpha_j| )}{2|k|}}\right) = C_1(M(k)/k).
\ee
To solve \eqref{inteq}  we iterate  the equation  to obtain
\begin{align*}
X(x) &= X_f \\ 
&+\sum_{j=1}^{m}(-1)^j  \int_{t_{j-1}}^{x_{f}}\cdots \int_{t_1}^{x_{f}}\int_{x}^{x_{f}}\frac{\mu^j V(t_1)\cdots V(t_{m-1})}{(2ik)^j}A(t_1,k)\cdots
A(t_j,k)X_fdt_1\cdots dt_j+ E_m,\\
\end{align*}
where
$$
E_m=(-1)^{m+1}\int_{t_{m}}^{x_{f}}\cdots  \int_{t_1}^{x_{f}} \int_{x}^{x_{f}}\frac{\mu^{m+1}V(t_1)\cdots V(t_{m+1})}{(2ik)^{m+1}}A(t_1,k)\cdots
A(t_m,k)X(t_{m+1})dt_1\cdots dt_{m+1}.
$$
The matrix $A(x,k)$ has rank one and can be written $A(x,k)=a(x,k)b(x,k)^T$ where $a(x,k)=\twomat{e^{-ikx}}{-e^{ikx}}$ and $b(x,k)=\twomat{e^{ikx}}{e^{-ikx}}$. Using that
$b(t_i,k)^Ta(t_{i+1},k)=2i\sin((t_i-t_{i+1})k)$ we can write
$$
\frac{1}{(2ik)^{m+1}}A(t_1,k)\cdots
A(t_m,k) = 
\frac{1}{(2ik)}a(t_1,k)\frac{\sin((t_1-t_{2})k)}{k}\cdots\frac{\sin((t_{m-1}-t_{m})k)}{k}
b(t_m,k)^T.
$$
so that
$$
\left|\frac{1}{(2ik)^{m+1}}A(t_1,k)\cdots A(t_m,k)\right|
\le \frac{1}{2|k|}C_2(k)^m
$$
where $C_2(k)$ is bounded for $|k|$ in a bounded set.
This bound together with $\|a(x,k)\|\|b(x,k)\|\le M(k)$ leads to
\begin{align*}
\|E_m\| &\le \frac{\mu^{m+1}}{2|k|} \int_{t_{m}}^{x_{f}}\cdots  \int_{0}^{x_{f}}|V(t_1)|\cdots |V(t_{m+1})| C_2(k)^m M(k) C_1(M(k)/|k|)dt_1\cdots dt_{m+1}\\
&\le \frac{\mu^{m+1}}{2|k|(m+1)!}\|V\|_1^{m+1} C_2(k)^m C_1(M(k)/|k|)
\end{align*}
{where $\|V\|_1 = \|V_0\|_1 + \sum_j |\alpha_j|$}.  This shows that the series for $X$ converges for all $\mu$. 
Proposition \ref{f2smallmu} follows from computing the first few terms in this series.


\vfill\eject
\end{document}